\documentstyle[emulateapj]{article}
\def\Msun{{\,M_\odot}}
\def\Zsun{{\,Z_\odot}}

\slugcomment{{\it Accepted by the Astrophysical Journal, 16 June 2003}}
\begin{document}

\title{The Star-Forming Dwarf Galaxy Populations of two $z \sim$ 0.4 Clusters:
MS1512.4+3647 and Abell 851}

\author{Jennifer M. Lotz \altaffilmark{1,2}, Crystal L. Martin \altaffilmark{3,4},
Henry C. Ferguson \altaffilmark{5}}

\altaffiltext{1}{Department of Physics and Astronomy,
Johns Hopkins University, Baltimore, MD 21218}
\altaffiltext{2}{Santa Cruz Institute for Particle Physics, University of California, Santa Cruz, CA 95064; jlotz@scipp.ucsc.edu}
\altaffiltext{3}{Department of Physics, University of California, Santa Barbara, CA 93106; cmartin@physics.ucsb.edu}
\altaffiltext{4}{Visiting astronomer, Kitt Peak National Observatory, AURA}
\altaffiltext{5}{Space Telescope Science Institute, Baltimore, MD 21218; ferguson@stsci.edu}

\begin{abstract}
We present the results of a deep narrow-band [OII] 3727 \AA\ emission-line search
for faint ($g <$ 27), star-forming galaxies in the field of the $z=0.37$ MS1512.4+3647
cluster.   We find no evidence for an over-density of emission-line sources relative to
the field at $z \sim 0.4$ (Hogg et al. 1998), 
and therefore conclude that the MS1512.4+3647 sample is dominated 
by field [OII] emission-line galaxies which lie along the $\sim$ 180 Mpc line of sight 
immediately in front and behind the cluster.  This is surprising, given that
the previously surveyed $z=0.41$ cluster Abell 851 has 3-4 times the field emission-line 
galaxy density (Martin et al. 2000). 
We find that the MS1512.4+3647 sample is deficient in galaxies with intermediate
colors ($1.0 < g-i < 2.0$) and implied star-formation exponential 
decay timescales $\tau \sim$ 100 Myr - 1 Gyr that dominate the Abell 851 emission-line galaxy population.
Instead, the majority of [OII] emission-line galaxies surrounding the MS1512.4+3647 cluster
are blue ($g-i \leq 1.0$) and forming stars in bursts with $\tau < $ 100 Myr.  
In both samples, galaxies with the shortest star-formation timescales
are preferentially among the faintest star-forming objects. 
Their $i$ luminosities are consistent with young stellar populations 
$\sim 10^8 - 10^9 \Msun$, although an additional factor of ten in stellar mass could be hiding 
in underlying old stellar populations. We discuss the implications for the star-formation 
histories of dwarf galaxies in the field and rich clusters.

\end{abstract}

\section{INTRODUCTION}
The formation of stars in the smallest galaxies provides a strong test
for current theories for structure formation. Hierarchical models
require the production of many small halos to act as the building blocks 
for larger halos.  But star-formation within small dark matter halos
must be suppressed to a large degree to prevent the universe from ``over-cooling''
and the over-production of Local Group dwarf galaxies (Klypin et al 1999, 
Moore et al 1999).  This may be
achieved by invoking strong stellar feedback within dwarfs (e.g. Dekel \& Silk
1986) or by heating and photo-ionizing the gas in dwarf halos during the epoch of 
re-ionization (Bullock, Kravtsov, \& Weinberg 2000).  
However, neither simple wind models nor re-ionization alone
are able to reproduce the dependence of dwarf galaxy morphology
and star-formation history on its local environment.

Environment is perhaps the most important regulator of dwarf galaxy 
star-formation.  Gas-poor dwarf galaxies are the most strongly clustered
galaxies in the local universe, found only as the satellites of larger galaxies
or in clusters of galaxies (Binggeli, Tarenghi \& Sandage 1990, Hogg et al. 2003), 
and are dominated by old 
and intermediate age stellar populations (see reviews by Ferguson \& Binggeli 
1994, Grebel 1997).  Like late-type spirals and star-forming dwarf galaxies 
(dIs), dwarf ellipticals (dEs) and dwarf spheroidals (dSphs)
possess exponential surface brightness profiles. The existence of an 
intermediate-class of dwarfs with low gas fractions, low star-formation rates, 
significant rotation (Simien \& Prugniel 2002),  
and in several cases, residual spiral arm structure (Jerjen, Kalnajs, \& Binggeli 2000), 
imply an evolutionary transition between late-type gas-rich dwarfs and dEs (Lin \& Faber 1983,
Moore, Lake, \& Katz 1998).
However, many dEs do not rotate (Geha, Guhathakurta, \& van der Marel 2002), 
and have higher surface brightnesses (Bothun et al. 1986)
and  higher chemical abundances (Grebel, Gallagher, \& Harbeck 2003) than expected for dIs which have 
simply lost their gas and stopped forming stars. 

The origin of these dEs remains unknown. 
Several different possibilities imply distinct star-formation and accretion 
histories for the cluster dE populations. If dwarf formation is strongly suppressed 
after re-ionization, dwarfs in high density regions may be more able to 
survive because they are more likely to collapse and form stars before the epoch of 
re-ionization (Tully et al. 2001). 
The pressure of the intra-cluster gaseous medium can prevent strong winds
from blowing out much of the dwarfs' gas and allow 
them to quickly form multiple generations of stars (Babul \& Rees 1992). 
In this scenario, the dE cluster populations are among the oldest cluster members. 
On the other hand, dynamical studies of the dwarf galaxy 
populations in the nearby Virgo and Fornax clusters suggest that the dE populations 
are not as relaxed as the giant elliptical populations and may have been recently accreted by
the cluster potentials (Drinkwater, Gregg, \& Colless 2001; 
Conselice, Gallagher, \& Wyse 2001).  Over time, processes such as ram-pressure stripping, 
strong tidal interactions, and ``galaxy harassment'' may remove the gas and halt star-formation 
in galaxies in cluster environments as they are accreted from the field (e.g. Moore, Lake \& Katz 1998).
Compared to the field populations at similar epochs, bright cluster galaxies have
suppressed star-formation rates and lower fractions of star-forming and blue galaxies
(Balogh et al. 1997, Ellingson et al 2001).  Gas removal by the cluster environment
may be even more efficient for low-mass galaxies and could transform star-forming dwarfs 
into dEs and dSphs. Therefore, many dEs may be recent cluster acquisitions with relatively 
young stellar populations.

At $z \geq$ 0.4,  where H$\alpha$ is shifted out of the optical wavelength 
region,  the [OII] 3727 \AA\ emission doublet is one of strongest 
optical tracers of the low density, photo-ionized nebulae 
associated with star-forming regions.  Early high resolution spectroscopy by Koo et al. (1997) 
identified several strong [OII] 3727 \AA\ emission-line dwarf galaxies in the $z \sim 0.4$ 
cluster Cl 0024+1624.  In Paper 1 (Martin, Lotz \& Ferguson 2000), we conducted  
a deep narrow-band [OII] 3727 \AA\ emission-line search for the faint star-forming
galaxies in the extremely massive and morphologically irregular cluster Abell 851 at $z= 0.41$. 
We detected several hundred [OII] emission-line galaxy candidates with
a number density 3-4 times that observed for field [OII] galaxies at similar redshifts. 
However, we found a deficit of extremely high [OII] equivalent width objects relative to the field
and concluded that Abell 851's environment is suppressing strong star-formation. Therefore,
many of Abell 851's faint star-forming galaxies could fade into the dE cluster population observed today.

In this paper, we present a similar [OII] 3727 \AA\ emission-line search in the less massive
but more relaxed $z=0.37$ cluster MS1512.4+3647 and compare the properties of 
its star-forming dwarf population to that of Abell 851.
In \S 2, we present the narrow-band [OII] 3727 \AA\ observations of MS1512.4+3647, 
and identify interlopers by their broad-band colors.
In \S 3, we compare the luminosity function and clustering properties of
MS1512.4+3647 [OII] emission-line candidates
to the Abell 851 and field [OII] emission-line populations.
We find that, unlike Abell 851, the MS1512.4+3647 [OII] luminosity function
shows no excess above the field, and is likely dominated by field
[OII] emitters which surround the cluster.  
In \S 4, we constrain star-formation histories and timescales of
the field-dominated MS1512.4+3647 star-forming galaxy sample and compare the derived 
star-formation histories to those found for Abell 851.  
We discuss the impact of the two cluster environments on the 
star-forming galaxy population at $z \sim 0.4$ and the implications for the formation of dEs. 
We assume $H_o$ = 70 km/s, $\Omega_m$ = 0.3, $\Omega_{\Lambda}$ = 0.7 
throughout this paper.

\section{EXPERIMENTAL DESIGN}

\subsection{MS1512.4+3647 Observations}
MS1512.4+3647 is a rich cluster at $z =  0.372$, first identified by the 
Einstein Medium Sensitivity Survey (Gioia et al. 1990, Stocke et al.1991)
with $L_X$~($2-10$ keV) $\sim 5.6~\times~10^{43}$~erg~s$^{-1}$.  This cluster has been the
subject of much study because of the detection of the lensed proto-galaxy
candidate cB58 at $z=2.7$ close to the central cluster galaxy (Yee et al. 1996)
and is part of the CNOC1 cluster sample (Canadian Network for Observational Cosmology Cluster
Redshift Survey; Abraham et al. 1998). 
The cluster appears to be elongated along the line 
of sight with a complex velocity structure and velocity dispersion $\sigma$ 
$=$ 575 km s$^{-1}$ (Borgani et al. 1999). High resolution ROSAT X-ray maps of 
MS1512.4+3647 are relatively smooth and elliptical, and the X-ray emission
is centered on the central cluster galaxy (Lewis et al. 1999).  
MS1512.4+3647's  virial mass and radius are approximately $8~\times
10^{14}~\Msun$ and 2~Mpc (Molikawa et al. 1999).

MS1512.4+3647 was observed June 20-23 1998 at the
 Kitt Peak National Observatory
4-meter telescope at prime focus using the T2KB CCD.  The on-band filter
W021 ($\lambda_c = 5135$ \AA, $\delta\lambda = 100.35$ \AA) was chosen to
match the redshifted [OII] 3727 \AA\ emission doublet of the cluster galaxies.
The off-band filter W022 ($\lambda_c = 5261$ \AA, $\delta\lambda = 44.24$ \AA)
samples the continuum 126 \AA\ redward of the on-band.
Emission-line galaxies with velocities within $+ 10 \sigma$ and $- 5 \sigma$ 
of the mean cluster velocity are detectable with this filter setup. 
The width of the on-band filter allows us to observe 
[OII] 3727 \AA\ emission-line objects along a line of sight much deeper ($\sim$
180 Mpc) than the angular width of our field of view ($\sim 4 \times 4$ Mpc at $z=0.37$).
Deep broad-band images in the SDSS $u$ ($\lambda_c = 3513$ \AA), $g$ 
($\lambda_c = 4759$ \AA), and $i$ ($\lambda_c = 7734$ \AA) bandpasses
were taken as well during the third night.

The images were processed using the IRAF CCD reduction package CCDPROC. 
Each image was over-scan subtracted and bias subtracted.  The images were then
corrected for variation in pixel sensitivity using dome flats for each
filter. Twilight sky images were used to correct for any 
residual large-scale illumination patterns.
The cluster images were aligned to within 0.1 pixels, combined and cosmic-ray
corrected with the iterative IRAF cosmic-ray rejection algorithm CRREJ.
Although observing conditions were photometric throughout the run,
the W021 and W022 images taken during the fourth night had lowered sensitivity
because the CCD had a slightly higher temperature that night.  
These images were scaled to
match the flux level of previous nights' images before they were combined.
The final images subtend $\sim$ 12.5~\arcmin\ $\times$ 12.7~\arcmin\ with typical seeing
 $\sim$ 1.4~\arcsec\ in W021, W022, g, and i and $\sim$ 2.2~\arcsec\ in u.
The astrometric solution was calculated with the IRAF task IMCOORDS.CCMAP using
the positions of  $\sim$ 50 stars from the Guide Star Catalog II  
\footnote{The Guide Star Catalogue-II is a joint project of the Space
                 Telescope Science Institute and the Osservatorio Astronomico di
                 Torino. Space Telescope Science Institute is operated by the
                 Association of Universities for Research in Astronomy, for the
                 National Aeronautics and Space Administration under contract
                 NAS5-26555. The participation of the Osservatorio Astronomico di
                 Torino is supported by the Italian Council for Research in
                 Astronomy. Additional support is provided by European Southern
                 Observatory, Space Telescope European Coordinating Facility, the
                 International GEMINI project and the European Space Agency
                 Astrophysics Division.}.
The resulting solution is accurate to $\pm$ 0.4 \arcsec\ in RA and $\pm$ 0.8 \arcsec\ in DEC.

A number of photometric and spectro-photometric standard stars 
(Landolt 1992; Massey et al. 1988)
were observed throughout the run and used to calibrate the flux of
the broad-band and narrow-band cluster images.  The $UBVRI$ photometry of the 
Landolt (1992) standards were converted to  SDSS $ugi$ magnitudes using 
the Fukugita et al. (1996) transformations. 
Archived HST WFPC2 F814W images of the central region of 
MS1512.4+3647 were also used to calibrate the $i$ band image.  The uncertainty
in the broad-band calibration is 0.15 magnitudes in i, 0.09 in g, and 0.12
in u.  The uncertainty in the narrow-band calibration is 7 \% in both bands.
The final cluster images have 4 $\sigma$ limiting magnitudes $\sim$
 25.5 in $u$, 27.3 in $g$, and
25.8 in $i$.  The limiting on and off band flux densities are  $1.1 \times 10^{-19}$
and $1.3 \times 10^{-19}$ erg s$^{-1}$ cm$^2$ \AA$^{-1}$. The [OII] 3727 \AA\ line flux 
(excess on-band flux) is $\sim 2.5 \times 10^{-17}$ erg s$^{-1}$ cm$^2$  at the
30 \% completeness level, which
equals a star-formation rate (SFR) of 0.13 $\Msun$ yr$^{-1}$ at the distance of the cluster
(see \S 4.2.1 for [OII] to SFR conversion assumptions).

\subsection{Detection of the faintest emission-line objects}
 
The sum of the final on-band and off-band images was used
as the detection image for the galaxy photometry software package SExtractor
version 1.2 (Bertin \& Arnouts 1996).
This summed image was convolved with a 4 pixel FWHM Gaussian and searched
for objects with 3 or more joined pixels with a flux level $\geq$ 1.5 $\sigma$ 
above the sky noise.  A total of 2883 objects was detected in the convolved 
summed image.  The isophotal apertures determined from the summed image were used to measure
the object fluxes in the W021, W022 and broad-band images. 
Using a model of the image's point spread function, 
SExtractor attempts to distinguish stars from galaxies by their isophotal area and flux
and assigns each object a probability that it is a star. 
Objects with SExtractor ``stellarity index''\ 
CLASS\_STAR $>$ 90\%  were considered stars and removed from the catalog (Figure 1).  
This left 2633 remaining ``galaxies''.  Confusion between stars and galaxies is 
inevitable at fluxes $< 5 \times 10^{-18}$ erg s$^{-1}$ cm$^{-2}$ \AA$^{-1}$, 
but galaxies should greatly outnumber stars at those flux levels.  
We expect to observe $\sim$ 200 stars fainter than $5 \times 10^{-18}$ erg s$^{-1}$ 
cm$^{-2}$ \AA$^{-1}$, assuming the Galaxy model B in Reid et al. (1996).  We found 12 objects
classified as stars by SExtractor with on-band excess emission $>$  4 $\sigma$.
However, all of these objects have continuum fluxes brighter than 2 $\times 10^{-17}$
erg s$^{-1}$ cm$^{-2}$ \AA$^{-1}$ and are too bright to be mis-classified emission-line
galaxies.

The [OII]  3727 \AA\ line flux was calculated assuming 
\begin{equation}
T_{ON} F_{[OII]} = F_{ON} - F_{OFF} \frac{\int^{ON} T(\lambda) d\lambda}
{ \int^{OFF} T(\lambda) d\lambda}
\end{equation}
where $F_{ON}$ and $F_{OFF}$ are the observed on- and off-band fluxes and
$T_{ON}$ is the average transmission of the on band filter.  
The error in the line flux depends on the error in both the on- and off-band
fluxes, but is dominated by the error in the continuum level determination
because the off-band filter W022 is less than half as wide as the on-band filter W021.
While a large number of excess on-band objects
were detected, only 95 objects have line fluxes greater than 4 $\sigma$
(Figure 2).  A number of excess off-band objects were detected as well;
we believe that many of these may be galaxies with [OII] emission or a 
strong 4000\AA\ break associated with a known sub-clump behind the 
cluster at $z = 0.41$ (Abraham et al. 1998).

The detection of emission-line sources and the measurement of 
their excess on-band flux become increasingly
difficult as the sources decrease in surface brightness and line flux. Galaxies
which are intrinsically very faint in continuum but have strong [OII] emission
may be detected only in our on-band image.  In this case, the uncertainty in the
continuum flux will dominate the error in the line flux determination
and only objects with high equivalent widths (EW) will have significantly 
detected on-band excesses.   On the other hand, the
weakest emission-line galaxies may have bright continua and 
be well detected in both the on and off
band images, but the emission-line flux may be on the level of a few times 
the uncertainty in the on-band flux.  
Thus our emission-line search will be incomplete for both intrinsically faint 
objects with low continuum fluxes and for objects with higher continuum fluxes
but faint [OII] emission.  Because the sum of the on- and off-band 
images was used to detect the objects, our ability to detect an object
is primarily a function of its summed on- and off-band flux and will decrease
as the object's total summed flux approaches the detection limit.

However, the detection of a galaxy depends not only on 
its total flux but its surface brightness as well. Galaxies with normally
observable total fluxes but large scalelengths and low surface brightnesses
may not be detected because the flux per pixel is comparable to the sky noise.
We do not find objects in the summed image above the 1.5 $\sigma$ SExtractor 
detection level with central surface brightness $\mu_0$ fainter than 1.0
 $\times 10^{-17}$ erg s$^{-1}$ cm$^{-2}$ arcsec$^{-2}$ ( = 26.9 AB magnitudes 
arcsec$^{-2}$).  Simulations of artificial galaxies 
with fixed total flux and varying surface brightness were added to the original image 
to determine the effect of surface brightness on the incompleteness at a given total flux.  We
find that the detection efficiency drops rapidly at $\mu_0$ $>$ 26.5 magnitudes 
arcsec$^{-2}$ for total fluxes fainter than $\sim 10^{-16}$ erg s$^{-1}$ cm$^{-2}$.  

To compute an effective incompleteness correction that accounts for the scatter  
in $\mu_0$ at each total flux level, we simulated galaxies with the same distribution
of observed surface brightnesses as the original image.  We chose real galaxies within
a given flux range at random from the SExtractor output catalog.  The total flux,
semi-major axis length, and ellipticity of each chosen galaxy were used to model that
galaxy using the IRAF ARTDATA.MKOBJECTS task. The simulated galaxy was added at random
positions to the original image 100 times.  SExtractor was run again on the modified image
to determine the detection efficiency for that galaxy.  This process was repeated 
for 10 randomly selected galaxies within each 0.1 dex flux bin, and the resulting 
detection efficiencies were averaged to compute the effective completeness for that flux bin.
We find that the effective completeness for galaxies with $\mu_0 <$ 26.9 magnitudes 
arcsec$^{-2}$ is  85\% for galaxies with total summed fluxes brighter than 
$8 \times 10^{-17}$ erg s$^{-1}$ cm$^{-2}$ and 30\% for galaxies fainter than 
$3 \times 10^{-17}$ erg s$^{-1}$ cm$^{-2}$.

\subsection{Interlopers/False detections}

We find 95 objects with excess on-band flux greater than 4 $\sigma$
in the $\sim$ 160 arcmin$^2$ cluster field.  A fraction of these detections
will not be [OII] 3727 \AA\ emission-line galaxies associated with the cluster, 
but rather foreground or background galaxies with spectral features which 
create excess on-band emission.  Here we determine how many such false [OII]
detections have contaminated our sample and identify the
probable interlopers.

\subsubsection{Broad-band color selection}
To determine which galaxies may produce false [OII] 3727\AA\ detections, we
have calculated the excess on-band EW expected as a function of redshift 
for irregular, spiral, and elliptical galaxy spectral templates 
(Figure 3, Coleman, Wu, \& Weedman 1980). The 
[OIII] 5007 \AA\ and H$\beta$ emission of foreground irregular 
galaxies at $z \leq$ 0.03 and 0.06  
will show up as excess on-band objects.  However, the volume 
sampled for [OIII] and H$\beta$ by our on-band is only 14 Mpc$^{3}$ and 
67 Mpc$^3$ respectively.  
We expect to detect $\sim$ 3 foreground emission-line galaxies given
the local density of emission galaxies (Salzer 1989).

Elliptical galaxies in the cluster may also produce 
excess on-band flux. This is because the CN3883 \AA\ absorption band
found in elliptical galaxy spectra (Davidge \& Clark 1994) falls into our
off-band at $z \sim 0.37$.  The CN absorption band lowers our estimation of
the continuum and results in false on-band excesses 
with typical rest-frame equivalent widths (EW) $\leq$ 20 \AA.
For this paper, we will assume that galaxies with red colors and on-band excess 
flux are not true [OII] emitters.  It is possible that a fraction of these objects are 
forming stars, but follow-up spectroscopy is needed to confirm the presence
of [OII] emission.

The on-band will also probe a large volume at high redshifts. Background star-bursting
galaxies may have strong Lyman $\alpha$ emission which will fall into our on-band filter
at redshifts $3.18-3.26$.   Deep Lyman $\alpha$ surveys at $z$= 4.5 (Rhoads et al. 2000,
Hu, Cowie \& McMahon 1998) find $\sim$ 4,000 objects per square degree per 
unit redshift with EW$_o$ $>$ 80 \AA\ and line+continuum fluxes $>$  2.6 $\times$
10$^{-17}$ erg cm$^{-2}$ s$^{-1}$.  Therefore we expect $\sim$ 13 Lyman $\alpha$ emitters 
in our field of view sampled by our on-band.  However,  the surface density of
Lyman $\alpha$ emitters can vary by a factor of 1.5 from field to field 
(Rhoads et al. 2000) and are highly clustered (Steidel et al. 1999). 
Given the intrinsic field to field
variations and that our survey is slightly less sensitive than the
Rhoads et al. survey, we may detect few background Lyman $\alpha$ galaxies.

We use the broad-band colors in order to distinguish between star-forming
cluster galaxies and the elliptical and high redshift interlopers.
Figure 4  shows the broad-band color evolution of each spectral 
type with redshift.  Lyman $\alpha$ emitters will drop out of the $u$ image
and be easily distinguished from cluster members 
by their very red $u-i$ and $u-g$ colors.  Elliptical galaxies in the
cluster will also have red colors ($u-i \geq 3.0$, $g-i \geq 2.0$, $u-g 
\geq 1.5$).  Irregular star-forming galaxies will have very blue colors
($u-i \leq 1.5$, $g-i \leq 1.0$, $u-g \leq 0.5$), and spiral galaxies
will have intermediate colors.  We expect the majority of our true [OII]
3727 \AA\ emitters to have rest-frame colors similar to irregular and
spiral galaxies.  On-band excess objects with $g-i >$ 2.0 and $u-i >$ 3.0 
are probably interlopers and are removed from our sample.

\subsubsection{Spectroscopic surveys of MS1512.4+3647}
We have cross-checked our on-band excess detections with the results of the 
CNOC1 spectroscopic survey of MS1512.4+3647 (Abraham et al. 1998) to test 
the reliability of our emission-line detection and broad-band color
classification.  Abraham et al. (1998) obtained spectroscopy 
for $> 50\%$ of galaxies brighter than $r'$= 21 in $\sim 220\ \sq \arcmin$ 
surrounding MS1512.4+3647. They determined each observed galaxy's 
redshift and spectral type by cross-correlating its spectrum with template SEDs.  
Our observations include 36 cluster member galaxies and 63 
field galaxies targeted by the Abraham et al (1998) spectroscopic survey. 
We compare these galaxies' CNOC spectral
classification (E/S0, Spiral, or Emission-line/Im) to  
our broad-band colors and on-band fluxes (Figure 5, Table 1).
We find that our broad-band colors reliably predict the spectral type of the
galaxy, and allow us to eliminate elliptical galaxies from our emission-line
sample.  Once these objects are removed from our sample, our false detection 
rate is $<$  4\%.  We observe [OII] emission above the 4 $\sigma$ level
in over half of the spectroscopically classified late-type spirals and irregular galaxies.

We have assumed that most on-band excess galaxies redder than $g-i = 2.0$ are not
[OII] emitters, but elliptical galaxies with strong absorption at CN3883 \AA. 
Six of these red on-band excess galaxies have been observed spectroscopically by 
Ziegler \& Bender (1997; Table 2).  The [OII] 3727\AA\ line did not fall into 
Ziegler \& Bender (1997) observed spectral range (6500 - 7500 \AA) and therefore we cannot directly
determine if these objects are truly [OII] emitters; however the H$\beta$ line, another
good star-formation indicator, was measured by Ziegler \& Bender (1997). 
The central cluster galaxy (2435, M09) has  H$\beta$ emission and is probably a true 
[OII] emission-line galaxy. The other five galaxies
do not have either strong H$\beta$ emission or absorption indicative of recent
star-formation, and are most likely false [OII] detections.
By removing the red excess on-band objects with $g-i$ $>$ 2.0 from our sample,
we may miss few true emission galaxies, but we will eliminate the largest source 
of contamination from the [OII] emission-line galaxy sample.

\subsubsection{ Lyman $\alpha$ candidates}
We expect that background Lyman $\alpha$ candidates will possess high EWs 
and be undetected in $u$.  We find three on-band excess objects with
EW$_o$ $>$ 80 \AA\ which drop out of our $u$ band image.  Object 1696
has an EW$_o$ $\sim$ 91 \AA , $i$ = 24.78 $\pm$ 0.13, and $g-i$ = 0.50 $\pm$ 0.16, 
consistent with a faint star-forming dwarf galaxy at $z = 0.37$.  However, the
other two high EW $u$-band drop-outs have EW$_o >> 100$ \AA\ (Table 3).
The rest-frame [OII] EWs of local star-forming galaxies rarely exceed 100 \AA, 
whereas Lyman $\alpha$ emitters could have rest-frame EW as high as
200 \AA\ (Charlot \& Fall 1993) and therefore EW$_o$ $\sim$ 800 \AA.
Object 2218 has an EW$_o$ $\sim$ 340 \AA, $g-i$ = 4.21 $\pm$ 0.21, is
undetected in $u$, and is therefore a good Lyman $\alpha$ candidate.
We also find one extremely high EW$_o$ ($\sim$ 1100 \AA) faint object (413) 
which is not detected in $u$ and may be a high redshift Lyman $\alpha$ emitter as well. 
This object is detected in the on-band and $g$ ( = 27.25 $\pm$ 0.36), but not in $i$.  
Assuming $g-i \sim$ 1.5 for Lyman $\alpha$ galaxies at $z= 3.2$, the expected $i$ 
magnitude would be 25.8, just at our 4 $\sigma$ detection limit.  
We cannot rule out the possibility that object 413 is a dwarf 
undergoing massive starburst with a rest-frame [OII] EW $\sim$ 800 \AA,
but it is more likely that it is background Lyman $\alpha$ emitter and
 we exclude it from our [OII] emission-line candidates.
We do not find any other Lyman $\alpha$ candidates above the 4$\sigma$ cutoff.
Gravitational lensing by the cluster potential is expected to increase, not decrease, the
number of detected Lyman $\alpha$ galaxies as the effect of luminosity amplification factor
should dominate over angular scattering by the lens on the observed high redshift background
galaxy number counts (Broadhurst, Taylor \& Peacock 1995).  A lensing model 
would be needed to determine the significance of the low number of Lyman $\alpha$
candidates towards the cluster core.  However, the low number we find is not obviously
in conflict with the populations described by Rhoads et al. (2000) and Hu et al. (1998).

\subsection{MS1512.4+3647 [OII] emission-line candidates}
We have detected 66 [OII] emission-line candidates in the MS1512.4+3647
field with excess on-band fluxes greater than 4 $\sigma$
and $g-i$ $<$ 2.0. In Table 4, we give the positions, colors, and fluxes 
of the emission-line candidates. 
In Figure 6, we plot the observed [OII] integrated line flux against the
 continuum flux for all detections above the 3 $\sigma$
level.  The red interlopers are plotted as triangles, and the 
galaxies with spiral and irregular colors are plotted as circles and stars
respectively. Objects with $g-i$ photometric errors $>$ 0.5 are plotted as 
open squares. The vertical solid lines are lines of constant equivalent width.
Contours of constant summed flux show where the data are 85\% and 30\% 
complete.   The upper dashed line is the 4 $\sigma$ cutoff, 
below which objects are excluded from our sample (see Figure 2), 
and the lower dashed line is the 3 $\sigma$ cutoff. 
All of the 4 $\sigma$ [OII] EW $> 100$ \AA\  emission-line
candidates were visually inspected and found to be spatially extended
and significantly detected in at least one broad-band image, 
therefore they are unlikely to be spurious noise detections.
We do not detect dwarfs above the 4 $\sigma$ cutoff with star-formation 
rates below 0.13 $\Msun$ yr$^{-1}$ (T$_{ON}$F[OII] $< 2.5 \times 10^{-17}$ erg s$^{-1}$ cm$^{-2}$).

\section{CLUSTER MEMBERS OR FIELD GALAXIES?}
The question of cluster membership for the MS1512.4+3647
 emission-line objects is a key issue that must be addressed. 
We expect very few foreground [OIII] 5007 \AA\ emitters and, with our color 
selection, we are confident that the majority of our detections 
are [OII] 3727 \AA\ emitters at $z \sim$ 0.37 and not background 
Lyman $\alpha$ galaxies.  
However, in addition to star-forming galaxies 
bound to the cluster, our on-band filter samples field [OII] 3727 \AA\ 
emission-line galaxies that lie along the $\sim$ 180 Mpc line of sight immediately
in front and behind the MS1512.4+3647 cluster. In order to determine
the field contamination of our MS1512.4+3647 [OII] candidates,  we compare their density 
and clustering properties to the field [OII] galaxies at $z \sim$ 0.4, as well 
as the $z = 0.41$ cluster Abell 851.  

\subsection{[OII] 3727 \AA\ Luminosity Function}
We find that the MS1512.4+3647 [OII] 3727 \AA\ luminosity 
function shows little evidence for an excess star-forming galaxy population 
associated with the cluster. We have calculated the [OII] luminosities 
of the MS1512.4+3647 emission-line candidates
from their integrated line fluxes assuming a spherical symmetry 
and correcting the observed line flux for the average transmission of our on-band, 
$T_{ON}$ = 72\%.  At L[OII] = 2.4 $\times 10^{40}$ erg s$^{-1}$ 
(T$_{ON}$ F[OII] = 5.0  $\times 10^{-17}$ erg s$^{-1}$ cm$^{-2}$), 
the MS1512.4+3647 luminosity function is $\sim 85\%$ complete down to 
EW $\sim$ 10 \AA\ (Figure 6).   We used the simulations described in 
section 2.2 to determine the probability $p_{ij}$ of detecting a galaxy with
continuum flux $i$ and line flux $j$.  The number of [OII] emission-line 
candidates is multiplied by the number of 
missing galaxies per detected galaxy, $1/N \sum^N 1/p_{ij}$.  
The incompleteness-corrected number counts were divided by the 
co-moving volume sampled by the on-band filter in our 
12.5\arcmin\ $\times$ 12.7\arcmin\ field of view (2802 Mpc$^3$).
The resulting luminosity function is given in Table 5.

In Figure 7, we compare the MS1512.4+3647 [OII] 3727 \AA\ 
luminosity function to the $z \sim 0.4$ field
[OII] luminosity function from the spectroscopic survey of Hogg et al. (1998),
the $z=0.41$ Abell 851 cluster luminosity function 
\footnote{We have found a $+0.43$ magnitude error in the Abell 851 $g$ photometry 
given in Paper 1, due to an error in the previously published Abell 851 $g'$ photometry
(Dressler \& Gunn 1992) used for our original calibration.  
In this paper, we have re-calibrated our Abell 851 $g$ photometry with new high-quality 
$BVRI$ photometry (I. Smail, private communication) and the SDSS-Johnson photometry
conversions computed by Fukugita et al. (1996).   We have re-selected the Abell 851 [OII] 
emission-line candidates based on the revised $g-i$ colors 
($g-i_{new} < 2.0$, $g-i_{new} = g-i_{old} - 0.43$).  
This has increased the Abell 851 [OII] emission-line candidates by 42 objects.}
and the $z=0.37$ foreground field of Abell 851 from Paper 1 (Martin et al. 2000).
The Hogg et al. (1998) spectroscopic survey is 90\% 
complete to 23 $R$ magnitude at [OII] EW $\sim$ 10 \AA, 
$\sim$ 1 magnitude deeper than our survey limit at
EW = 10 \AA.  The Abell 851 survey was also a narrow-band imaging
survey using a slightly different filter setup, and is complete down to
EW $\sim$ 10 \AA\ at L[OII] $= 10^{40}$~erg~s$^{-1}$ (see Figure 5 in Paper 1).
The foreground field of A851 was observed with the same filter setup and similar
exposure times as MS1512.4+3647, and its selection effects are similar to those for
our MS1512.4+3647 sample.

The [OII] luminosity function of MS1512.4+3647 shows little excess relative 
to the field at $z \sim 0.4$, even at bright [OII] luminosities where 
our survey is complete.  This suggests that the majority
 of our detections are simply field galaxies immediately in front and behind the 
cluster for which [OII] 3727\AA\  falls into our on-band and the density
of star-forming galaxies bound to MS1512+3647 is indistinguishable from the field.  
On the other hand, Abell 851 luminosity function shows an excess of 
[OII] emission-line galaxies 3-4 times the density
of the field [OII] galaxies at $z=0.4$ (Paper 1) and MS1512.4+3647 sample.  

\subsection{Clustering properties of [OII] emission-line candidates}
We find that the MS1512.4+3647 emission-line objects are not strongly
clustered towards the central cluster galaxy (filled square, top of 
Figure 8). This is in contrast to
the galaxies with elliptical-like colors ($g-i > 2.0$, open squares),   
which are more likely to be found around the central cluster galaxy at 
projected radii $<$ R$_{200}$. ( R$_{200}$, the radius at which the mean inner density of the cluster
is 200 times the critical density of the universe, $\sim$ 1.2 Mpc for MS1512.4+3647 
assuming cluster velocity dispersion $\sigma$ = 575 km s$^{-1}$.) 
At $z=0.37$, our field of view is $\sim$ 4 Mpc $\times$ 
4 Mpc and  will include most of the cluster.  
The surface density of [OII] emitters is consistent with the
field [OII] emission-line population (Hogg et al. 1998).
Except for the inner 100 \arcsec\ ($\sim 770$ kpc),  
Abell 851 star-forming galaxies are also weakly clustered relative to
the red galaxies (bottom of figure 8) out to 1.5 R$_{200}$, 
even though the surface density of star-forming galaxies 
observed in the Abell 851 field is well above that expected for 
field [OII] emitters. 

Given that the MS1512.4+3647 luminosity function shows no over-density of 
[OII] emitters and the surface density of star-forming galaxies is likely to be dominated
by field galaxies at all cluster radii, it seems likely that most of 
our MS1512.4+3647 sample are field galaxies surrounding MS1512.4+3647 
which are not bound to the cluster potential.
MS1512.4+3647 does possess some galaxies which have had recent star-formation $-$
the CNOC1 survey found that $\sim$ 20 \% of  spectroscopically-confirmed 
cluster members brighter than M$_{r'} = -19.0$ in the core of MS1512.4+3647 are bluer than 
$g'-r' = 0.25$, like other clusters observed by CNOC1 at similar redshifts (Ellingson et al. 2001).  
If MS1512.4+3647 is a typical $z \sim 0.4$ cluster, then most clusters 
at $z \sim 0.4$  do not have high densities of star-forming galaxies, but are instead 
slowing swallowing up the field galaxy population.  Assuming accreted galaxies 
eventually cease star-production and have detectable [OII] emission 
for $\sim$ 1 Gyr (Balogh et al. 2000), then 
a average $z \sim$ 0.4 cluster has assembled most of its 
mass at redshifts $\geq$ 0.5.  

Abell 851, on the other hand, is clearly an unusual cluster in terms of its
mass, X-ray luminosity, and density of star-forming galaxies. 
Its virial mass is approximately 10 times that of MS1512.4+3647 and its bolometric X-ray
luminosity is over twice that of MS1512.4+3647 (16.08 $\times 10^{-44}$
erg s$^{-1}$ vs. 7.62 $\times 10^{-44}$ erg s$^{-1}$;  Wu et al. 1999). 
This cluster has significant sub-structure in its X-ray emission (Schindler \& 
Wambsganss 1996) and galaxy distribution (Kodama et al. 2001). 
It appears to be a cluster in formation,
and has most likely acquired its large number of star-forming galaxies from
a collapsing system of groups and filaments as opposed to gradual accretion of the
surrounding field population.

The vast difference between the MS1512.4+3647 and Abell 851 
[OII] emission-line galaxy densities implies that the density of star-forming galaxies 
in a cluster relative to the field may be an strong tracer of the recent 
assembly history of the cluster.   Balogh et al. (2002) reached a 
similar conclusion when they found a significantly higher density of H$\alpha$
emitting galaxies in the $z=0.18$ cluster Abell 1689 than in the more relaxed but
more distant cluster AC 114 at $z=0.31$.  Normalizing each of our cluster's densities of 
[OII] emitters by their bolometric X-ray luminosities and correcting for selection effects,
we find that the normalized density of star-forming galaxies above
the MS1512.4+3647 detections limits is 0.0058 $\pm$ 0.0005 Mpc$^{-3}$ 
L$_X$( 10$^{44}$ erg s$^{-1}$)$^{-1}$ for Abell 851 and  0.0031 $\pm$ 
0.0004 Mpc$^{-3}$ L$_X$( 10$^{44}$ erg s$^{-1}$)$^{-1}$. 
These values are a lower limit for Abell 851 as our field of view encompassed
only the inner half of the cluster (Paper 1), and a upper limit for MS1512.4+3647 
due to the significant field contamination. Therefore Abell 851 has at least twice
(and possibly ten times) as many star-forming galaxies per unit volume and 
X-ray luminosity than the more typical cluster MS1512.4+3647.  

\section{STAR-FORMATION PROPERTIES AND HISTORIES OF EMISSION-LINE GALAXIES}

In this section, we use the [OII] emission-line equivalent widths and broad-band
colors of the MS1512.4+3647 [OII] candidates to constrain their
recent star-formation.  We then compare the derived
star-formation histories of the MS1512.4+3647 field-dominated and the Abell 851 
cluster-dominated [OII] emission-line galaxy samples 
for clues to the origin of the disparity in the density of star-forming 
galaxies between the two clusters.   Finally, we discuss the implications for 
dwarf galaxy evolution in the field and clusters.

\subsection{[OII] 3727 \AA\ emission as a tracer of star-formation}

The dependence of [OII] luminosity
on the star-formation rate has been empirically calibrated using both
H$\alpha$ and H$\beta$ emission (Kennicutt 1992; Gallagher, Bushouse \& Hunter 1989).
We adopt the observed [OII] 3727 \AA\ to H$\beta$ relation F[OII] = 3.2 F$_{H\beta}$
derived for local star-forming galaxies (Gallagher et al. 1989).  
To convert the observed (and dust-extincted) [OII] flux to a star formation rate, we 
assume a dust-correction for the observed H$\beta$ flux, derive the intrinsic
ratio of H$\beta$ to H$\alpha$ flux, and assume an intrinsic H$\alpha$ to star formation rate
relation.    We adopt $A_B = 1.0$ extinction correction,  
typical of nearby star-forming dwarfs and irregular galaxies
(Hunter \& Hoffman 1999).  Assuming nebular temperature
$=10^4$, Case B recombination, and  $A_{H\alpha}$ = 0.59 for $A_B = 1.0$,  
we derive extinction-corrected $H\alpha$ fluxes 2.0 times larger than the 
measured, uncorrected [OII] fluxes. Finally, we adopt  Kennicutt, Tamblyn, \& Congdon's (1994)  
intrinsic H$\alpha$-SFR calibration, assuming the K83 initial
mass function (Kennicutt 1983) from 0.1 to 100 $\Msun$ :
\begin{equation}
SFR (\Msun\ yr^{-1}) = \frac{L_{H\alpha}}{ 1.36 \times 10^{41} {\rm erg\ s^{-1}}}.
\end{equation}

By choosing an extinction correction appropriate for low-mass galaxies, 
we will systematically underestimate the extinction (and SFR) in luminous 
galaxies by a factor of 2-3 (Kennicutt 1992) or more. The observed [OII]/H$\alpha$ 
ratio depends strongly on reddening and metallicity and can vary by an order of magnitude
for galaxies with $-14 > $M$_B$ $> -22$ (Jansen, Franx, \& Fabricant 2001).
Comparison of star-formation rates derived
from [OII]~3727~\AA\ emission and near-infrared, far-infrared, and 
radio data of ``post-starburst'' cluster galaxies at $z \sim$ 0.4 and 
local dusty starburst galaxies suggests that optical extinction 
could be as high as A$_B = 2-3$ magnitudes (Poggianti
\& Wu 1999, Smail et al. 1999). But these extreme extinction levels 
are found primarily in the most massive ``E+A'' galaxies and dusty starbursts.
Dwarf galaxies have low metallicities
and thus likely to have much less dust than these massive galaxies.
Thus, while we may under-estimate the 
extinction and intrinsic star-formation rates in  massive, dusty
and metal-rich galaxies in our sample,
our assumption of modest extinction and a relatively low [OII]/H$\alpha$ 
ratio should be appropriate for the majority of our detections.

\subsection{MS1512.4+3647 [OII] emission-line galaxy star-formation histories}

The [OII] emission-line population detected in the field immediately surrounding 
MS1512.4+3647 is dominated by blue ($g-i < 1.0$), faint ($i > 21$) galaxies.
At $z \sim 0.4$, the $u-i$ and $g-i$ colors span the redshifted 4000 \AA\ break 
and are therefore sensitive to the luminosity-weighted age of the observed stellar population.  
In Figure 9 and 10, we compare the broad-band luminosities and colors of the MS1512.43647
4 $\sigma$ on-band excess detections to population synthesis model tracks 
for an instantaneous burst and constant star-formation 
with constant stellar mass and varying age (Bruzual \& Charlot 2000).
The broad-band colors of the  MS1512.4+3647 [OII] emission-line candidates 
imply a young ($<$ 100 Myr) fading burst or an somewhat older ($\sim$ 500 Myr)
continuous star-formation history.  The models assume a metallicity $\sim 1/3 \Zsun$
and Calzetti (1997) dust extinction with $A_B = 1.0$; 
adopting a higher metallicity or extinction would give younger ages.
The red CN 3883\AA\ absorption-line interlopers clearly separate from the bluer 
[OII] emission-line candidates in the color-color diagrams.  These bright red low-EW 
interlopers have colors and luminosities consistent with an unreddened 1-5 Gyr, 
10$^{10}$ - 10$^{11}$ $\Msun$ burst.

The majority of MS1512.4+3647 [OII] emission-line candidates are dwarf
galaxies.  Comparison of the galaxies' colors and $i$ magnitudes to the 
population synthesis models allow us to place a lower limit on
their stellar masses (Figure 10).   Many of the 
emission-line candidate have implied stellar masses $\sim$ 10$^{8} - 10^{9} \Msun$. 
However, any underlying population of old stars will be virtually undetectable, 
and up to 90\% of a galaxy's stellar mass could be hidden 
in a $\sim 10$ Gyr population by a 40 Myr burst without significantly reddening its $g-i$ and $u-i$ colors.   
Nevertheless, even with an additional factor of ten in stellar mass, 
many of the faint star-forming galaxy would still have sub-L$^{*}$ stellar masses ($< 10^{10} \Msun$).
The faint [OII] emission-line galaxies tend to be compact in our $i$ images 
with projected radii less than a few kpc, giving further evidence that
these galaxies are intrinsically small.  Archived high-resolution 
HST WFPC2 F814W images of MS1512.4+3647 contain only three [OII] emission-line
candidates from our sample.  One is the central galaxy (2435), which has an 
elliptical halo but bright irregular nucleus and several close companions; 
the other two objects appear to be distorted spirals (2565 and 2437). 

The MS1512.4+3647 objects with the highest EWs are 
also the faintest and bluest galaxies (Figure 10 and 11).  
Locally, galaxies with rest frame [OII] EW $>$ 40 \AA\ tend to
be very late-type low-mass Sdm/Im galaxies (Kennicutt 1992) and many of
the blue dwarfs observed by Gallagher et al. (1989) have [OII] EWs $\sim$ 50 \AA\
or greater.  Therefore many of these faint, high EWs objects are 
probably late-type, low-mass galaxies undergoing rapid star-formation.  
In Figure 11, we compare the colors and observed [OII] EWs
to the A$_B = 1.0$ reddened and redshifted population synthesis model 
tracks for three different exponentially decaying star-formation timescales 
(SFR $\propto$ exp($-t/\tau$), $\tau$ = 10 Myr, 100 Myr, and 1 Gyr) 
and a constant star-formation model  ($\tau = \infty$). 
The model [OII] EWs were calculated  from their instantaneous SFRs using the 
empirical conversion described in \S4.2.1. Older, continuously star-forming galaxies have
relatively high EWs due to the constant production of new stars but 
redder continua due to the older stellar population.  Young fading bursts 
have rapidly decreasing EWs as the massive stars die off but
remain relatively blue as long as [OII] emission is detectable.
We find that the majority of the MS1512.4+3647 [OII] candidates' colors and [OII] EWs
are best described by an exponentially decaying model with $\tau \leq$ 
100 Myr and ages 20-300 Myr. 

\subsection{MS1512.4+3647 v. Abell 851}
The distribution of colors and [OII] EWs for the MS1512.4+3647 [OII] emission-line
galaxy candidates appears more like that of the faint field population at 
$z \sim 0.4$ than the rich cluster Abell 851. 
Abell 851 possesses a large number of [OII] emission-line candidates with 
intermediate $g-i$ colors and [OII] EWs consistent with star-formation decay 
timescales $\tau$ $\sim$ 100 Myr - 1 Gyr \footnote{The recalibration of our Abell 851 $g$ photometry
has strengthened this conclusion, as the revised colors have been shifted bluewards
and allow more intermediate-color galaxies to meet our [OII] emission-line candidate criteria.} 
(Figure 12; see also Paper 1).  Such galaxies are rare in 
the foreground Abell 851 field (Figure 12) and the MS1512.4+3647 cluster field.
We also find fewer high [OII] EW objects in the Abell 851 cluster sample than in
the MS1512.4+3647 sample or the foreground Abell 851 field.
In Abell 851, we found only 7 objects with EW $>$ 100 \AA\ compared to 
39 high EW off-band ``field'' sources.   Direct comparison to our MS1512.4+3647 
sample is difficult, as many of the detected Abell 851 on- and off-band high EW objects 
have line fluxes well below the MS1512.4+3647 limiting line flux.  Nevertheless, 
we do find over twice as many faint EW $>$ 100 objects with F[OII] $> 10^{-17}$
erg s$^{-1}$ cm$^{-2}$ in our MS1512.4+3647 3 $\sigma$ detections 
as we do for the Abell 851 cluster and it is probable that more high EW objects lie 
below MS1512.4+3647 3 $\sigma$ detection limit (Figure 6) .  
  
In Paper 1, we concluded that the lack of high [OII] EW Abell 851 cluster galaxies
was evidence for the suppression of starbursts in Abell 851 relative to the field 
galaxy population. Infalling galaxies are predicted to have their gaseous reservoirs 
removed by the cluster environment on timescales of 1-3 Gyr 
(Balogh et al. 2000). If star-formation in field galaxies is gradually suppressed 
as they enter the cluster environment, one may expect to find a population of 
cluster galaxies with slowly fading star-formation. 
The deficit of intermediate color, $\tau > $ 100 Myr objects in the MS1512.4+3647 
and the foreground field of Abell 851 implies that relatively few 
emission-line galaxies have been accreted and ``strangled'' by the MS1512.4+3647 
cluster within the previous few Gyrs.
 
The suppression of star-formation within the cluster core
 should be evident as a dependence on star-formation rate
(and [OII] emission) on cluster-centric radius.  In Figure 13, we plot the average 
SFR and L[OII] of the MS1512.4+3647 and Abell 851 [OII] emission-line candidates 
with projected radius. The average SFR of the central regions 
(r $<$ 100\arcsec) of both clusters is dominated by the 
central cluster galaxies, which appear to be undergoing minor mergers.  
Otherwise, the mean star-formation rate is approximately constant with 
projected radius out to $\sim$ 3 R$_{200}$ in MS1512.4+3647 and $\sim$ 1 R$_{200}$ in
Abell 851.   If the MS1512.4+3647 sample is dominated by field galaxies,
we would not expect to see any radial trend. Also, the effect of the cluster environment on
infalling galaxies is believed to be strongest outside of the virial radius 
(Ellingson et al. 2001, Diaferio et al. 2001) and therefore
outside of our field of view for Abell 851.  Therefore it is difficult to 
rule out galaxy ``strangulation'' in these two clusters, despite the 
lack of radial dependence on the observed average star-formation rate.

\subsection{Implications for dwarf galaxy evolution}

Most of the faint star-forming galaxies in the field 
surrounding MS1512.4+3647 are best described by low-mass star-bursts 
with $\tau <$ 100 Myr, as are the off-band excess objects in the field in
front of Abell 851 (Figure 11 and 12).  Galaxies with longer star-formation
timescales above our detection limit of 0.13 $\Msun$ yr$^{-1}$ are absent from 
our ``field'' galaxy samples, implying that short bursts dominate the 
star-formation in many field galaxies.  Both the derived star-formation histories of
local irregular galaxies (Grebel 1997) and the lack of red, faded, 
low-mass field galaxies in the Hubble Deep Field at $z < 0.5$ suggest that 
field dwarf galaxies (Lotz 2003; Ferguson \& Babul 1998)
undergo multiple bursts throughout their evolution.    

However, further star-formation may be suppressed if the galaxy falls into a rich 
cluster environment.  Local clusters are filled with thousands of dwarf
galaxies which are gas-poor and no longer forming stars.
If the observed [OII] emission-line galaxies fade quickly and 
do not undergo another episode of star-formation,  many could become as faint 
as present day dEs.  A galaxy with an $\tau$ = 10 Myr exponentially decaying burst, 
an age = 40 Myr, and an observed $i$ magnitude $\sim$ 22 
at $z=0.37$ would fade into a M$_R$ $\sim$ $-16$ by $z=0$.
A galaxy with $\tau$ = 100 Myr, age = 300 Myr, and an observed $i$ 
magnitude $\sim$ 22 at $z=0.37$ would fade into a M$_R$ $\sim$ -18 by $z=0$.

At $z \sim$ 0.4, we are only able to observe the progenitors of cluster dEs 
brighter than M$_R = -15.0$, which number $\sim$ 300 in Virgo.  To produce 
this number of faded emission-line galaxies in MS1512.4+3647,
over four times the number of [OII] candidates observed
in the MS1512.4+3647 field must accrete onto the cluster by the present day.  
Correcting the observed density of field emission-line galaxies
for the duty cycle, this requires a field volume infall rate greater than 50 Mpc$^{3}$ 
Gyr$^{-1}$ in order to accrete similar numbers of bright dEs in the 4-5 Gyr since $z= 0.4$.  
Our observations of MS1512.4+3647 suggests a maximum infall rate
for emission-line galaxies of a few Mpc$^{3}$ Gyr$^{-1}$, if recently accreted 
galaxies are able to form stars and have [OII] emission visible for $\sim$ 1 Gyr.  
Therefore it is unlikely that most bright dEs in typical clusters were acquired by the 
gradual infall of star-forming field galaxies since $z \sim 0.4$. 
But if most dE progenitors are accreted in clumps as groups merge intermittently 
with the cluster, as is the case for Abell 851,  then one or two such major 
mergers could account for the majority of the dE population observed in 
local clusters.  In this case, clusters which have assembled 
more recently would have significantly younger dE populations.

In the Virgo cluster, bright nucleated dE (dE,N) 
and faint dSph/dEs are as spatially clustered as the giant ellipticals 
(Ferguson \& Sandage 1989).  Could a $z \leq 0.4$ infalling galaxy population 
become as clustered by $z=0$ as local dE populations? 
Dynamical friction timescales grow longer with smaller satellite 
mass and  $\leq$ 10$^9 \Msun$ galaxies would sink quite slowly in 
the cluster potential $-$ dwarf galaxies accreted 4-5 Gyr ago do not
have enough time to become as spatially clustered as the local dE 
cluster population. However, the Fornax and Virgo dwarf galaxy 
populations (which are dominated by dE,N) have higher velocity dispersions 
than the giant elliptical galaxy populations (Drinkwater, Gregg \& 
Colless 2001; Conselice, Gallagher, \& Wyse 2001), which is consistent with recent infall. 
One way to reconcile the spatial distribution of dE,N with their high 
cluster velocities is to form the nuclei as the dwarfs pass through 
the cluster center on timescales less than a few Gyr.  Gas-rich dwarfs may
form centralized stellar excesses via galaxy harassment in a few Gyr and 
galaxies which form nuclei are more likely to survive within the central
regions of clusters than those that do not (Moore, Lake \& Katz 1998).  
Most dE nuclei are compact, globular cluster-like 
objects and could also be formed by the decay of massive globular clusters 
into the center via dynamical friction (Lotz et al. 2001).  The timescale for
this process may be shorter for dwarfs within the central regions of clusters
than those on the outskirts (Oh \& Lin 2000).

\section{SUMMARY}

We have presented the results of a deep narrow-band [OII] 3727\AA\ emission-line survey 
for faint star-forming galaxies in the $z = 0.37$ MS1512.4+3647 cluster.
Using broad-band $u-i$ and $g-i$ colors, we are able to distinguish $z \sim$ 0.37
[OII] emission-line candidates from $z \sim$ 0.37 elliptical galaxies which may produce
false on-band excesses and background $z \sim$ 3.2 Lyman $\alpha$ emitters.  
We find two Lyman $\alpha$ candidates with
EW $>$ 300 \AA\ and undetected $u$ fluxes.   We identify 66 [OII] emission-line
candidates in the MS1512.4+3647 field. 

The observed density of [OII] emission-line galaxies surrounding
MS1512.4+3647 is identical to the field population at 
$z \sim$ 0.4, and is most likely dominated by [OII]-emitting field galaxies
within $\sim$ 180 Mpc of the cluster along the line of sight.
This is in strong contrast to the previously studied $z=0.41$ cluster Abell 851, 
which has an [OII] emission line galaxy density at least 3-4 times that of the
field.   We find that Abell 851 has 2-10 times as many star-forming galaxies per 
unit volume and X-ray luminosity as MS1512.4+3647.
Abell 851 is a cluster in formation, with many merging sub-clumps and filaments,
unlike the more typical, relaxed MS1512.4+3647 cluster.  Thus 
the density of star-forming galaxies in a cluster relative to the field appears to be
a strong tracer of its recent assembly history.

The field-dominated MS1512.4+3647 and foreground Abell 851 field samples lack the galaxies with   
star-formation decay timescales $\tau \sim$ 100 Myr - 1 Gyr that dominate the Abell 851 cluster
emission-line galaxy population.  Such star-formation timescales are expected if 
star-formation is gradually suppressed as field galaxies are accreted by the cluster (Balogh et al. 2000).  
Therefore, the low density of [OII] emission line galaxies and 
the absence of intermediate-color emission-line galaxies in MS1512.4+3647 suggests that
relatively few star-forming galaxies have been 
accreted by the MS1512.4+3647 cluster in the previous Gyr.  On the other hand, 
the abundance of intermediate color emission-line galaxies,
and the absence of high [OII] EW galaxies in Abell 851 are consistent with the 
recent accretion of many of Abell 851's star-forming galaxies.

The field emission-line galaxy population surrounding MS1512.4+3647 
is forming stars in bursts with $\tau \sim$ 10-100 Myr.  
Galaxies with the highest [OII] EWs and bluest colors in both samples 
are typically among the lowest luminosity systems.  
These faint starbursts have young stellar populations with masses 
$\geq 10^8 - 10^9 \Msun$, but could hide up to a additional factor of ten in stellar
mass in an old faded stellar population. 
If the observed star-forming galaxies around MS1512.4+3647 and Abell 851 
fall into the clusters' potential wells and rapidly cease star production, they could fade 
to be as faint as dE in local clusters.  However, the large numbers of present-day 
cluster dEs require typical infall rates much greater than those implied by our observations
of MS1512.4+3647.  The large dE cluster populations can be acquired
only if most have been accreted before $z \sim$ 0.4 or 
in several major merger events (as in Abell 851). 

We would like to thank Ian Smail for access to his photometry of Abell 851.  JML 
thanks Joel Primack, Piero Madau, and the Santa Cruz Institute for Particle Physics 
at UCSC for support during the final stages of this paper. 
CLM gratefully acknowledges financial support from the Sherman Fairchild Foundation
through Caltech and from NASA through the Hubble Fellowship Program.

\clearpage

\begin{deluxetable}{rccrrrccc}
\tablecaption{CNOC MS1512.4+3647 sample \tablenotemark{1}}
\tablefontsize{\footnotesize}
\tablehead{ \colhead{CNOC ID } & \colhead{$z$} & \colhead{Type\tablenotemark{2}}
 & \colhead{ID} & \colhead{on-band EW} & \colhead{S/N \tablenotemark{3}} 
& \colhead{$i$}  & \colhead{$u-i$} & \colhead{$g-i$}  \\
\colhead{} &\colhead{} &\colhead{} &\colhead{} &\colhead{ (\AA)} &\colhead{}  }
\startdata
82    & 0.368     &    1 &	783 &	  9.2 &   2.2 &  20.64 $\pm$  0.08 &  5.29 $\pm$ 0.91 &   1.94 $\pm$  0.09 \\
158   & 0.370     &    1 &	879 &	  6.1 &   3.0 &  19.63 $\pm$  0.08 &  3.64 $\pm$ 0.16 &   1.75 $\pm$  0.09 \\
1819  & 0.368     &    1 &	946 &	  8.4 &   2.9 &  20.00 $\pm$  0.08 &  4.49 $\pm$ 0.29 &   2.00 $\pm$  0.09 \\
318   & 0.368     &    1 &     1834 &	 21.8 &   3.4 &  21.08 $\pm$  0.08 &   $>$ 4.4        &   1.65 $\pm$  0.09 \\
586   & 0.372     &    1 &     2021 &	  8.1 &   2.5 &  20.42 $\pm$  0.08 &  4.89 $\pm$  0.50 &   1.98 $\pm$  0.09 \\
600   & 0.370     &    1 &     2023 &	  9.4 &   2.8 &  20.21 $\pm$  0.08 &  5.34 $\pm$  0.70 &   2.03 $\pm$  0.09 \\
643   & 0.373     &    1 &     2065 &	 11.8 &   9.6 &  18.13 $\pm$  0.08 &  5.19 $\pm$ 0.21 &   2.22 $\pm$  0.09 \\
815   & 0.363     &    1 &     2174 &	  4.3 &   1.5 &  20.18 $\pm$  0.08 &  6.60 $\pm$ 2.13 &   1.93 $\pm$  0.09 \\
1290  & 0.371     &    1 &     2554 &	 14.5 &   5.7 &  19.60 $\pm$  0.08 &  4.17 $\pm$ 0.18 &   2.26 $\pm$  0.09 \\
1397  & 0.372     &    1 &     2636 &	 13.9 &   2.5 &  20.99 $\pm$  0.08 &  4.65 $\pm$ 0.59 &   2.08 $\pm$  0.09 \\
1534  & 0.372     &    1 &     2736 &	 20.5 &   3.0 &  21.23 $\pm$  0.08 &  $>$ 3.2  	      &   2.20 $\pm$  0.09 \\
147   & 0.369     &    2 &	825 &	  9.9 &   1.5 &  21.20 $\pm$  0.08 &  6.09 $\pm$ 2.58 &   2.00 $\pm$  0.09 \\
2080  & 0.376     &    2 &     1132 &	 -3.2 &   1.1 &  20.36 $\pm$  0.08 &  3.98 $\pm$ 0.23 &   1.97 $\pm$  0.09 \\
2377  & 0.371     &    2 &     1285 &	  8.8 &   3.3 &  19.75 $\pm$  0.08 &  5.67 $\pm$ 0.68 &   2.16 $\pm$  0.09 \\
355   & 0.382     &    2 &     1880 &	 15.9 &   3.7 &  20.65 $\pm$  0.08 &  4.86 $\pm$ 0.57 &   2.01 $\pm$  0.09 \\
508   & 0.371     &    2 &     1965 &	  0.7 &   0.3 &  19.47 $\pm$  0.08 &  3.92 $\pm$ 0.16 &   2.06 $\pm$  0.09 \\
486   & 0.374     &    2 &     1995 &	 15.8 &   4.8 &  19.95 $\pm$  0.08 &  5.21 $\pm$ 0.49 &   2.16 $\pm$  0.09 \\
613   & 0.369     &    2 &     2041 &	 22.5 &   8.0 &  19.69 $\pm$  0.08 &  3.97 $\pm$ 0.16 &   2.21 $\pm$  0.09 \\
710   & 0.370     &    2 &     2101 &	  8.9 &   4.1 &  19.43 $\pm$  0.08 &  4.00 $\pm$ 0.17 &   1.97 $\pm$  0.09 \\
1164  & 0.374     &    2 &     2453 &	 18.3 &   6.6 &  19.58 $\pm$  0.08 &  5.21 $\pm$ 0.37 &   2.44 $\pm$  0.09 \\
972   & 0.369     &    2 &     2463 &	-11.4 &   1.8 &  21.62 $\pm$  0.08 &  $>$ 3.8         &   2.21 $\pm$  0.09  \\
1401  & 0.373     &    2 &     2631 &	 27.3 &   6.0 &  20.36 $\pm$  0.08 &  5.02 $\pm$ 0.51 &   2.38 $\pm$  0.09 \\
1698  & 0.366     &    2 &     2861 &	 25.4 &   9.1 &  19.60 $\pm$  0.08 &  3.89 $\pm$ 0.17 &   2.07 $\pm$  0.09 \\
396   & 0.366     &    4 &     1944 &	  7.8 &   7.6 &  19.48 $\pm$  0.08 &  2.00 $\pm$ 0.11 &   1.24 $\pm$  0.09 \\
377   & 0.370     &    4 &     1952 &	-11.4 &   2.7 &  21.11 $\pm$  0.08 &  3.19 $\pm$ 0.22 &   1.82 $\pm$  0.09 \\
817   & 0.376     &    4 &     2177 &	  2.8 &   1.2 &  20.08 $\pm$  0.08 &  2.81 $\pm$ 0.12 &   1.77 $\pm$  0.09 \\
824   & 0.377     &    4 &     2350 &	 11.1 &   5.8 &  20.22 $\pm$  0.08 &  1.52 $\pm$ 0.11 &   0.94 $\pm$  0.09 \\
1250  & 0.368     &    4 &     2495 &	  6.1 &   1.2 &  21.43 $\pm$  0.08 &  3.15 $\pm$ 0.24 &   1.58 $\pm$  0.09 \\
142   & 0.379     &    5 &	830 &	 18.6 &   9.2 &  20.74 $\pm$  0.08 &  1.40 $\pm$ 0.11 &   0.95 $\pm$  0.09 \\
161   & 0.378     &    5 &	846 &	 18.7 &   5.1 &  21.36 $\pm$  0.08 &  2.11 $\pm$ 0.14 &   1.16 $\pm$  0.09 \\
191   & 0.378     &    5 &	873 &	  6.8 &   3.8 &  20.53 $\pm$  0.08 &  1.60 $\pm$ 0.11 &   1.01 $\pm$  0.09 \\
2063  & 0.381     &    5 &     1095 &	  5.9 &   2.9 &  20.42 $\pm$  0.08 &  1.74 $\pm$ 0.11 &   1.18 $\pm$  0.09 \\
2166  & 0.372     &    5 &     1187 &	 25.2 &  15.0 &  20.06 $\pm$  0.08 &  1.39 $\pm$ 0.11 &   0.91 $\pm$  0.09 \\
370   & 0.365     &    5 &     1882 &	 -1.0 &   0.9 &  19.52 $\pm$  0.08 &  1.99 $\pm$ 0.11 &   1.31 $\pm$  0.09 \\
1094  & 0.372     &    5 &     2435 &	 18.3 &  21.7 &  17.72 $\pm$  0.08 &  3.12 $\pm$ 0.11 &   1.74 $\pm$  0.09 \\
769   & 0.379     &    5 &     2128 &	  8.7 &   2.6 &  21.37 $\pm$  0.08 &  2.22 $\pm$ 0.15 &   1.13 $\pm$  0.09 \\
\enddata
\tablenotetext{1}{Abraham et al. 1996}
\tablenotetext{2}{ 1,2 = E/S0, 4= Sbc, 5=Emission-line/Im}
\tablenotetext{3}{S/N of the on-band excess flux}
\end{deluxetable}

\clearpage
\begin{deluxetable}{ccccccccrccc}
\tablecolumns{12}
\tablewidth{0pc}
\tablecaption{ZB97 MS1512.4+3647 Elliptical Galaxy sample \tablenotemark{1}}
\tablefontsize{\footnotesize}
\tablehead{ \colhead{Z\&B ID} & \colhead{$z$} & \colhead{CNOC ID} & \colhead{$I$}
& \colhead{$V-I$} & \colhead{H$\beta$ EW}
 & \colhead{ID} & \colhead{on-band EW} &\colhead{S/N} & \colhead{$i$}  
& \colhead{$g-i$} & \colhead{$u-i$}  \\
&\colhead{} & \colhead{} & \colhead{} & \colhead{}  & \colhead{(\AA)} & \colhead{} & \colhead{(\AA)} 
}
\startdata
M02  &  0.372 &     \nodata  &   19.40  &  2.34 &    1.83 &    2280  &   18.9 &   6.4 & 	 19.79 $\pm$ 0.08 &   2.30 $\pm$  0.09  &  4.79 $\pm$ 0.29 \\
M09  &  0.372 &     1094  &   18.70  &  2.32 &   -1.07 &   2435 & 18.3 &  21.7 &  17.72 $\pm$  0.08  &   1.74 $\pm$  0.09 &  3.12 $\pm$ 0.11\\ 
M11  &  0.373 &     \nodata  &   18.97  &  2.41 &    1.49 &    2440  &   23.8 &   9.8 & 	 19.13 $\pm$ 0.08 &   2.16 $\pm$  0.08  &  4.33 $\pm$ 0.20 \\
M15  &  0.373 &     643  &   18.29  &  2.18 &    1.69 &    2065  &   11.8 &   9.6 & 	 18.13 $\pm$ 0.08 &   2.22 $\pm$  0.09  &  5.19 $\pm$ 0.21 \\  
M17  &  0.364 &     \nodata  &   18.77  &  2.17 &    1.74 &    2346  &   17.9 &  11.3 & 	 18.92 $\pm$ 0.08 &   2.16 $\pm$  0.09  &  4.19 $\pm$ 0.15 \\
M19  &  0.368 &     710  &   19.31  &  2.18 &    1.66 &    2101  &   11.8 &   9.6 & 	 19.43 $\pm$ 0.08 &   1.97 $\pm$  0.09  &  4.00 $\pm$ 0.17  \\
\enddata
\tablenotetext{1}{Ziegler \& Bender 1997}
\end{deluxetable}

\begin{deluxetable}{rcccccc}
\tablecolumns{8}
\tablewidth{0pc}
\tablecaption{Background Lyman $\alpha$ emission-line candidates}
\tablefontsize{\footnotesize}
\tablehead{ \colhead{ID} &\colhead{RA \tablenotemark{1}}
 &\colhead{DEC} &\colhead{on-band flux} &\colhead{F$_{\lambda}$} &\colhead{$i$}
&\colhead{$g-i$} \\
\colhead{} &\colhead{J2000} 
&\colhead{J2000}
&\colhead{erg s$^{-1}$ cm$^{-2}$} &\colhead{erg s$^{-1}$ cm$^{-2}$ \AA$^{-1}$} 
&\colhead{mag} &\colhead{mag}} 
\startdata
413  & 15 14 01.55 &+36 31 56.4  &2.07 $\pm$ 0.51E-17  & 1.90E-20  &27.84 $\pm$ 1.07 &$-0.59 \pm 1.13$ \\
2218 & 15 14 23.70  & +36 35 41.3 &4.11 $\pm$ 0.72E-17  & 1.21E-19 &21.94 $\pm$ 0.08 & 4.21 $\pm$ 0.21 \\
\enddata
\tablenotetext{1}{RA $\pm$ 0.4 \arcsec, DEC $\pm$ 0.8 \arcsec}
\end{deluxetable}

\clearpage
\begin{deluxetable}{rccccccc}
\tablecolumns{8}
\tablewidth{0pc}
\tablecaption{MS1512.4+3647 [OII] emission-line candidates}
\tablefontsize{\footnotesize}
\tablehead{ \colhead{Id} &\colhead{RA \tablenotemark{1}}
 &\colhead{DEC} &\colhead{T$_{ON}$ F[OII]} &\colhead{F$_{\lambda}$} &\colhead{$i$}
&\colhead{$g-i$} &\colhead{$u-i$} \\
\colhead{} &\colhead{J2000} 
&\colhead{J2000}
&\colhead{(erg s$^{-1}$ cm$^{-2}$)} &\colhead{(erg s$^{-1}$ cm$^{-2}$ \AA$^{-1}$)} 
&\colhead{} &\colhead{} &\colhead{}} 
\startdata
\cutinhead{$g-i <$ 1.0}
  1696  &  15 14 10.55  & +36 42 17.4 & 3.25 $\pm$  0.81E-17 & 3.57E-19    &	24.78 $\pm$    0.13    &   0.50  $\pm$    0.16  &  $> -0.85$\\
  2825  &  15 14 40.16  & +36 38 09.5 & 4.7  $\pm$  1.2E-17  & 7.37E-19    &	23.05 $\pm$    0.09    &   0.90  $\pm$    0.10   &  1.56  $\pm$  0.20\\
   498  &  15 14 00.97  & +36 32 15.7 & 4.74 $\pm$  0.86E-17 & 2.70E-19    &	24.32 $\pm$    0.11    &   0.65  $\pm$    0.13  &  1.14  $\pm$  0.31\\
  2349  &  15 13 55.53  & +36 36 15.8 & 6.0 $\pm$   1.4E-17  & 2.33E-18    &	22.68 $\pm$	 0.08	 &   0.84  $\pm$    0.10   &  1.56  $\pm$  0.19\\
  2324  &  15 14 36.89  & +36 36 07.4 & 6.0 $\pm$   1.4E-17  & 2.03E-18    &	22.57 $\pm$	 0.08	 &   0.73  $\pm$    0.09   &  1.20  $\pm$ 0.14 \\
  1574  &  15 14 48.01  & +36 41 38.6 & 6.7 $\pm$   1.3E-17  & 1.22E-18    &	23.94 $\pm$	 0.11	 &  -0.17  $\pm$    0.12   &  1.19  $\pm$ 0.34\\
  2340  &  15 14 33.25  & +36 36 11.3 & 7.0 $\pm$   1.5E-17  & 2.62E-18    &	22.30 $\pm$	 0.08	  &  0.35  $\pm$    0.09   &  0.37  $\pm$ 0.11\\
  2565  &  15 14 20.31  & +36 37 01.1 & 7.0 $\pm$   1.5E-17  & 2.68E-18    &	22.24 $\pm$	 0.08	  &  0.54  $\pm$    0.09   &  0.70  $\pm$  0.12\\
   517  &  15 14 14.73  & +36 32 17.9 & 7.1 $\pm$   1.3E-17  & 1.05E-18    &	23.37 $\pm$	 0.09	  &  0.30  $\pm$    0.10   &  1.26  $\pm$  0.23\\
  1492  &  15 14 23.70  & +36 41 20.0 & 7.4 $\pm$   1.7E-17  & 2.34E-18    &	22.07 $\pm$	 0.08	  &  0.88  $\pm$    0.09   &  1.14  $\pm$  0.13\\
  2024  &  15 14 14.99  & +36 34 52.4 & 7.6 $\pm$   1.5E-17  & 2.23E-18    &	22.54 $\pm$	 0.08	  &  0.68  $\pm$    0.09   &  1.11  $\pm$  0.14\\
   135  &  15 14 11.03  & +36 30 39.1 & 7.7 $\pm$   1.5E-17  & 1.79E-18    &	22.48 $\pm$	 0.08	 &  0.97  $\pm$    0.09  &   2.44  $\pm$  0.31\\
   156  &  15 14 36.16  & +36 30 42.1 & 7.8 $\pm$   1.7E-17  & 1.81E-18    &	22.35 $\pm$	 0.08	 &  0.96  $\pm$    0.09  &   1.17  $\pm$  0.14\\
   616  &  15 14 40.46  & +36 32 40.7 & 8.3 $\pm$   1.5E-17  & 2.32E-18    &	22.29 $\pm$	 0.08	 &  0.96  $\pm$    0.09  &   1.82  $\pm$  0.18\\
  1075  &  15 14 43.57  & +36 39 19.7 & 9.9 $\pm$   1.6E-17  & 2.78E-18    &	22.22 $\pm$	 0.08	 &  0.66  $\pm$    0.09  &   1.43  $\pm$  0.14\\
  769   &  15 14 28.05  & +36 33 20.1 & 1.03 $\pm$  0.22E-16 & 1.49E-17  &    20.47 $\pm$     0.08    & 0.71  $\pm$    0.09  &   1.49  $\pm$  0.11 \\
 1107   &  15 14 23.70  & +36 39 26.8 & 1.08 $\pm$  0.21E-16 & 9.57E-18  &    21.16 $\pm$     0.08    & 0.28  $\pm$    0.09  &   0.40  $\pm$  0.11 \\
 2823   &  15 14 19.05  & +36 38 09.9 & 1.12 $\pm$  0.17E-16 & 3.88E-18  &    21.96 $\pm$     0.08    & 0.89  $\pm$    0.09  &   0.95  $\pm$  0.12 \\
 1909   &  15 14 17.14  & +36 34 21.2 & 1.16 $\pm$  0.18E-16 & 4.61E-18    &	21.69 $\pm$    0.08    &   0.79  $\pm$    0.09  &    1.41  $\pm$  0.13\\
 1305   &  15 14 43.43  & +36 40 27.5 & 1.20 $\pm$  0.18E-16 & 3.77E-18    &	21.61 $\pm$    0.08    &   0.79  $\pm$    0.09  &    1.74  $\pm$  0.13\\
  499   &  15 14 36.24  & +36 32 11.2 & 1.25 $\pm$  0.16E-16 & 3.45E-18    &	21.86 $\pm$    0.08    &   0.88  $\pm$    0.09  &    1.82  $\pm$  0.15\\
   72   &  15 14 39.55  & +36 30 20.3 & 1.44 $\pm$  0.16E-16 & 2.40E-18    &	22.33 $\pm$    0.08    &   0.90  $\pm$    0.09  &    1.67  $\pm$  0.18\\
 2037   &  15 14 08.57  & +36 34 55.9 & 1.51 $\pm$  0.18E-16 & 6.47E-18    &	21.63 $\pm$    0.08    &   0.49  $\pm$    0.09  &    1.34  $\pm$  0.12\\
 2633   &  15 14 13.61  & +36 37 17.2 & 1.54 $\pm$  0.19E-16 & 3.11E-18    &	22.05 $\pm$    0.08    &   0.57  $\pm$    0.09  &    1.08  $\pm$  0.13\\
  105   &  15 14 30.42  & +36 30 25.7 & 1.56 $\pm$  0.23E-16 & 9.57E-18    &	20.72 $\pm$    0.08    &   0.66  $\pm$    0.09  &    1.85  $\pm$  0.12\\
 2350   &  15 13 59.49  & +36 36 10.8 & 1.77 $\pm$  0.31E-16 & 1.60E-17    &	20.22 $\pm$    0.08    &   0.94  $\pm$    0.09  &    1.52  $\pm$  0.11\\
 1579   &  15 14 46.97  & +36 41 48.8 & 1.78 $\pm$  0.26E-16 & 3.39E-17    &	19.93 $\pm$    0.08    &   0.21  $\pm$    0.09  &    0.14  $\pm$  0.11\\
 1687   &  15 13 57.74  & +36 42 13.3 & 1.80 $\pm$  0.26E-16 & 2.21E-17    &	20.00 $\pm$    0.08    &   0.47  $\pm$    0.09  &    0.66  $\pm$  0.11\\
 1636   &  15 14 02.85  & +36 42 01.7 & 1.85 $\pm$  0.24E-16 & 3.59E-17    &	19.52 $\pm$    0.08    &   0.59  $\pm$    0.09  &    1.43  $\pm$  0.11\\
  830   &  15 14 21.35  & +36 33 36.7 & 1.87 $\pm$  0.20E-16 & 1.00E-17    &	20.74 $\pm$    0.08    &   0.95  $\pm$    0.09  &    1.40  $\pm$  0.11 \\
 1441   &  15 14 52.55  & +36 41 00.2 & 1.89 $\pm$  0.19E-16 & 6.27E-18    &	21.96 $\pm$    0.08    &  -0.27  $\pm$     0.09  &   0.09 $\pm$   0.11\\
 1438   &  15 14 16.63  & +36 41 03.7 & 1.99 $\pm$  0.17E-16 & 1.60E-18    &	22.97 $\pm$    0.09    &   0.40  $\pm$     0.10  &   3.79 $\pm$   1.81\\
  542   &  15 14 16.07  & +36 32 18.5 & 2.06 $\pm$  0.20E-16 & 6.88E-18    &	21.33 $\pm$    0.08    &   0.65  $\pm$     0.09  &   1.23 $\pm$   0.12\\
 2612   &  15 14 09.68  & +36 37 11.6 & 2.24 $\pm$  0.25E-16 & 1.30E-17    &	20.07 $\pm$    0.08    &   0.79  $\pm$     0.09  &   1.64 $\pm$   0.11\\
 1035   &  15 14 24.84  & +36 39 02.5 & 2.38 $\pm$  0.19E-16 & 5.68E-18    &	21.57 $\pm$    0.08    &   0.73  $\pm$     0.09  &   1.11 $\pm$   0.12\\
  949   &  15 14 24.01  & +36 38 43.5 & 2.82 $\pm$  0.19E-16 & 6.45E-18    &	21.37 $\pm$    0.08    &   0.75  $\pm$     0.09  &   1.02 $\pm$   0.11\\
  631   &  15 14 00.54  & +36 32 42.9 & 3.34 $\pm$  0.27E-16 & 1.85E-17    &	20.05 $\pm$    0.08    &   0.97  $\pm$     0.09  &   1.55 $\pm$   0.11\\
 1627   &  15 13 56.77  & +36 41 59.4 & 3.68 $\pm$  0.30E-16 & 4.71E-17    &	19.37 $\pm$    0.08    &   0.44  $\pm$     0.09  &   1.30 $\pm$   0.11\\
 1187   &  15 14 45.62  & +36 39 47.0 & 4.28 $\pm$  0.28E-16 & 1.70E-17    &	20.06 $\pm$    0.08    &   0.91  $\pm$     0.09  &   1.39 $\pm$   0.11\\
  160   &  15 14 09.22  & +36 30 40.3 & 4.94 $\pm$  0.33E-16 & 6.17E-17    &	18.81 $\pm$    0.08    &   0.72  $\pm$    0.09  &    1.62  $\pm$  0.11 \\
  596   &  15 13 55.50  & +36 32 29.8 & 5.04 $\pm$  0.30E-16 & 1.85E-17    &	20.15 $\pm$    0.08    &   0.81  $\pm$    0.09  &    1.46  $\pm$  0.11 \\
\cutinhead{1.0 $<$ $g-i$ $<$ 2.0} 
   992 & 15 14 25.52  & +36 38 57.2    &  4.19 $\pm$ 0.93E-17&  2.89E-19 & 24.13  $\pm$ 0.10 & 1.10 $\pm$  0.15 & 1.76 $\pm$ 0.47 \\
  1515 & 15 14 21.01  & +36 41 24.8    &  5.7  $\pm$  1.4E-17&  1.31E-18 & 22.69  $\pm$ 0.08 & 1.03 $\pm$  0.10 & 2.03 $\pm$ 0.25 \\
  1074 & 15 14 25.66  & +36 39 07.5    &  6.3  $\pm$  1.4E-17&  1.78E-18 & 21.69  $\pm$ 0.08 & 1.95 $\pm$  0.09 & 4.45 $\pm$ 0.88 \\
   416 & 15 14 49.79  & +36 31 50.7    &  6.3  $\pm$  1.4E-17&  1.31E-18 & 22.04  $\pm$ 0.08 & 1.52 $\pm$  0.09 & 2.25 $\pm$ 0.19 \\
  2136 & 15 14 15.03  & +36 35 21.9    &  6.4  $\pm$  1.1E-17&  7.62E-19 & 23.09  $\pm$ 0.09 & 1.06 $\pm$  0.10 & 2.10 $\pm$ 0.30 \\
  1013 & 15 14 10.70  & +36 39 02.0    &  7.3  $\pm$  1.4E-17&  1.66E-18 & 21.61  $\pm$ 0.08 & 1.74 $\pm$  0.09 & 1.99 $\pm$ 0.14 \\
  2164 & 15 14 24.69  & +36 35 28.2    &  7.7  $\pm$  1.4E-17&  1.68E-18 & 21.59  $\pm$ 0.08 & 1.72 $\pm$  0.09 & 2.43 $\pm$ 0.16 \\
  1951 & 15 14 21.39  & +36 34 33.5    &  8.7  $\pm$  2.0E-17&  4.80E-18 & 20.42  $\pm$ 0.08 & 1.93 $\pm$  0.09 & 4.25 $\pm$ 0.33 \\
   846 & 15 14 27.18  & +36 33 40.9    &  8.9  $\pm$  1.7E-17&  4.77E-18 & 21.36  $\pm$ 0.08 & 1.16 $\pm$  0.09 & 2.11 $\pm$ 0.14 \\
   186 & 15 14 33.30  & +36 30 50.1    &  1.05 $\pm$ 0.19E-16&  3.73E-18 & 20.96  $\pm$ 0.08 & 1.41 $\pm$  0.09 & 2.27 $\pm$ 0.13 \\
   159 & 15 14 53.18  & +36 30 41.4    &   1.06 $\pm$ 0.20E-16&  7.47E-18 & 21.08  $\pm$ 0.08 & 1.14 $\pm$  0.09 & 2.16 $\pm$ 0.14 \\
  1459 & 15 14 37.24  & +36 41 05.0    &   1.06 $\pm$ 0.23E-16&  1.21E-17 & 20.14  $\pm$ 0.08 & 1.37 $\pm$  0.09 & 2.58 $\pm$ 0.12 \\
  2101 & 15 14 16.41  & +36 35 11.7    &   1.11 $\pm$ 0.27E-16&  1.24E-17 & 19.43  $\pm$ 0.08 & 1.97 $\pm$  0.09 & 4.00 $\pm$ 0.17 \\
  1933 & 15 14 16.36  & +36 34 29.2    &   1.11 $\pm$ 0.26E-16&  1.26E-17 & 20.35  $\pm$ 0.08 & 1.05 $\pm$  0.09 & 1.86 $\pm$ 0.11 \\
  1384 & 15 14 44.04  & +36 40 45.2    &   1.14 $\pm$ 0.19E-16&  2.99E-18 & 21.81  $\pm$ 0.08 & 1.07 $\pm$  0.09 & 1.70 $\pm$ 0.15 \\
   724 & 15 14 03.09  & +36 33 10.2    &   1.17 $\pm$ 0.20E-16&  6.58E-18 & 21.21  $\pm$ 0.08 & 1.07 $\pm$  0.09 & 1.69 $\pm$ 0.12 \\
  1202 & 15 14 29.42  & +36 39 54.7    &   1.17 $\pm$ 0.25E-16&  1.56E-17 & 19.88  $\pm$ 0.08 & 1.46 $\pm$  0.09 & 2.88 $\pm$ 0.13 \\
  1599 & 15 14 27.09  & +36 41 51.9    &   1.40 $\pm$ 0.25E-16&  9.08E-18 & 19.74  $\pm$ 0.08 & 1.85 $\pm$  0.09 & 2.84 $\pm$ 0.12 \\
   821 & 15 14 28.32  & +36 33 31.3    &   1.46 $\pm$ 0.24E-16&  1.00E-17 & 20.36  $\pm$ 0.08 & 1.34 $\pm$  0.09 & 2.08 $\pm$ 0.12 \\
   120 & 15 14 30.60  & +36 30 30.0    &   1.48 $\pm$ 0.23E-16&  6.52E-18 & 20.04  $\pm$ 0.08 & 1.96 $\pm$  0.09 & 5.40 $\pm$ 0.72 \\
   672 & 15 14 03.32  & +36 32 51.0    &   1.85 $\pm$ 0.31E-16&  3.74E-17 & 19.25  $\pm$ 0.08 & 1.09 $\pm$  0.09 & 1.99 $\pm$ 0.11 \\
   749 & 15 14 18.43  & +36 33 13.5    &   1.94 $\pm$ 0.23E-16&  9.00E-18 & 20.65  $\pm$ 0.08 & 1.03 $\pm$  0.09 & 1.59 $\pm$ 0.11 \\
  2437 & 15 14 21.74  & +36 36 13.3    &   2.01 $\pm$ 0.29E-16&  7.76E-18 & 20.50  $\pm$ 0.08 & 1.38 $\pm$  0.09 & 2.14 $\pm$ 0.13 \\
  1944 & 15 13 58.95  & +36 34 32.5    &   2.13 $\pm$ 0.28E-16&  2.72E-17 & 19.48  $\pm$ 0.08 & 1.24 $\pm$  0.09 & 2.00 $\pm$ 0.11 \\
  1262 & 15 14 16.40  & +36 40 09.8    &   3.26 $\pm$ 0.38E-16&  4.50E-17 & 18.60  $\pm$ 0.08 & 1.61 $\pm$  0.09 & 3.44 $\pm$ 0.12 \\
  1042 & 15 14 11.64  & +36 39 07.4    &   3.27 $\pm$ 0.31E-16&  2.92E-17 & 19.41  $\pm$ 0.08 & 1.22 $\pm$  0.09 & 2.05 $\pm$ 0.11 \\
  2435 & 15 14 22.54  & +36 36 21.8    &  1.325 $\pm$ 0.06E-15&  7.23E-17 & 17.72  $\pm$ 0.08 & 1.74 $\pm$  0.09 & 3.12 $\pm$ 0.11 \\
\enddata
\tablenotetext{1}{RA $\pm$ 0.4 \arcsec, DEC $\pm$ 0.8 \arcsec}
\end{deluxetable}

\clearpage
\begin{deluxetable}{cccccc}
\tablecolumns{6}
\tablewidth{0pc}
\tablecaption{[OII] Luminosity Function of MS1512.4+3647 cluster field}
\tablehead{ \colhead{T$_{ON}$ F[OII]} & \colhead{L[OII]} & \colhead{N} 
&\colhead{$W$\tablenotemark{1}} &\colhead{N$_c$} &\colhead{$\Phi$} \\
\colhead{erg s$^{-1}$ cm$^{-2}$} &\colhead{erg s$^{-1}$} 
&\colhead{} &\colhead{} &\colhead{} &\colhead{Mpc$^{-3}$}}
\startdata
-15.0  & 41.68 & 1  & 1.00   & 1.00   & 0.00036 $\pm$ 0.00036 \\
-15.5  & 41.14 & 18 & 1.06   & 19.08  & 0.0068 $\pm$ 0.0016\\
-16.0  & 40.68 & 42 & 1.17   & 47.04  & 0.0168 $\pm$ 0.0026\\
\enddata
\tablenotetext{1}{Incompleteness correction factor}
\end{deluxetable}

\clearpage
\begin{figure}
\plotone{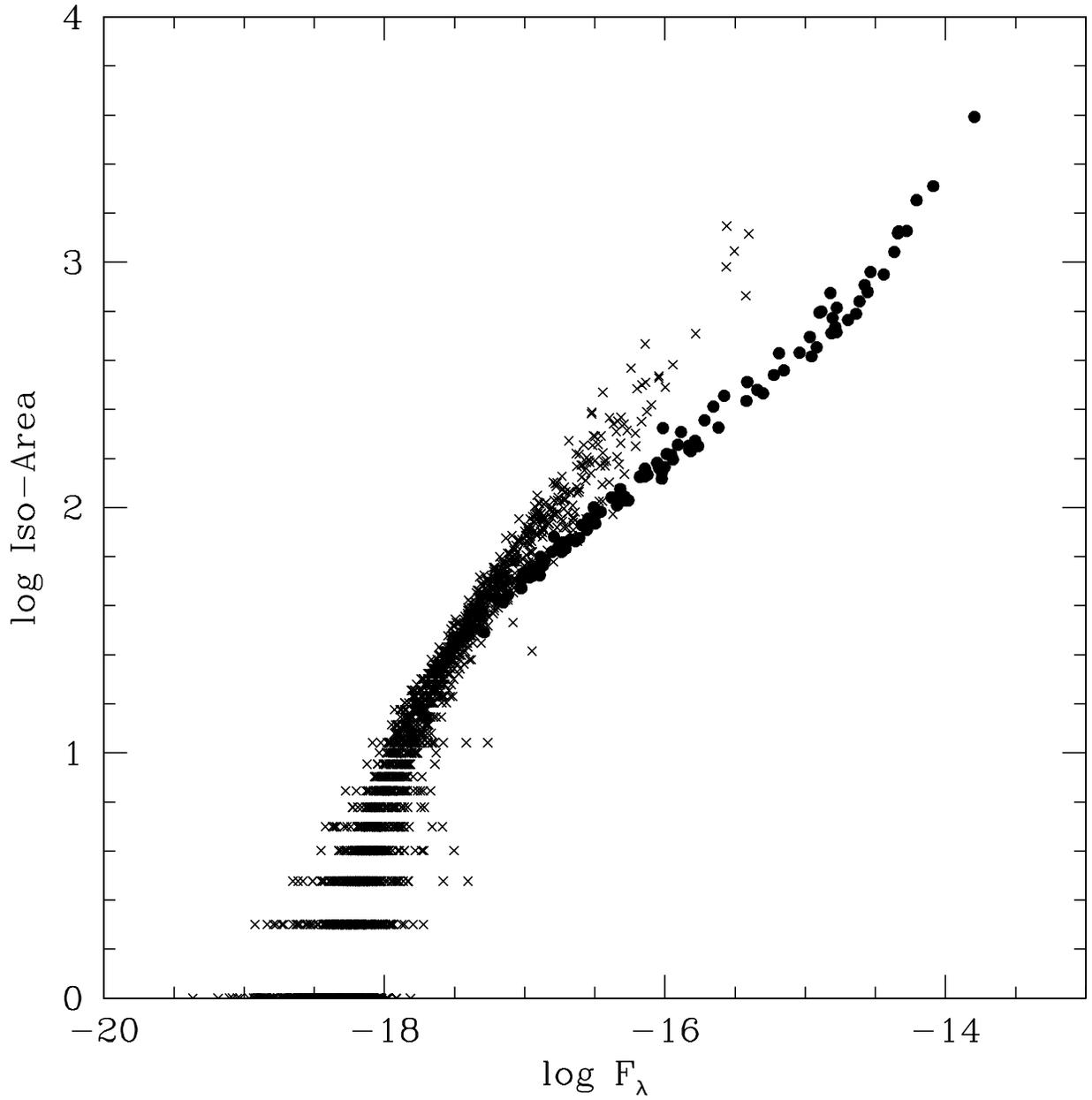}
\caption{log isophotal area (pixels) vs. log continuum flux F$_{\lambda}$ for
all detected objects. The filled circles have been classified as 
stars by their flux concentration.}
\end{figure}

\clearpage
\begin{figure}
\plotone{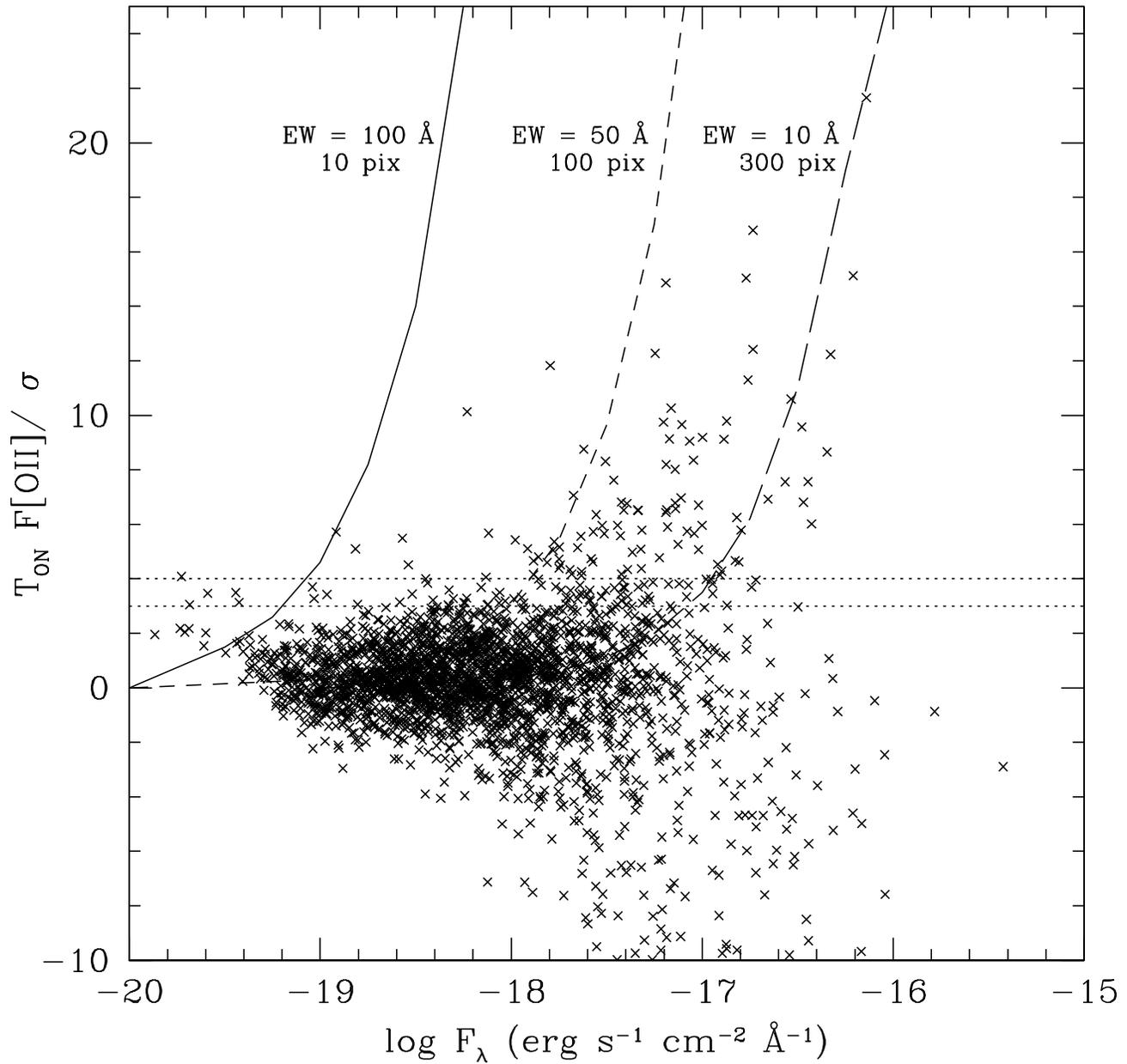}
\caption{The signal-to-noise ratio for F[OII] line flux vs. 
the continuum flux F$_{\lambda}$.  The 3 and 4 $\sigma$ cutoffs are marked
as dashed lines. The dependence of the survey's sensitivity on the [OII] EW 
and surface brightness is shown for three models:  [OII] EW 
of 100 \AA\ and isophotal area = 10 pixels; [OII] EW = 50 \AA\ 
and isophotal area = 100 pixels; and [OII] EW = 10 \AA\ 
and isophotal area = 300 pixels. }
\end{figure}

\clearpage
\begin{figure}
\plotone{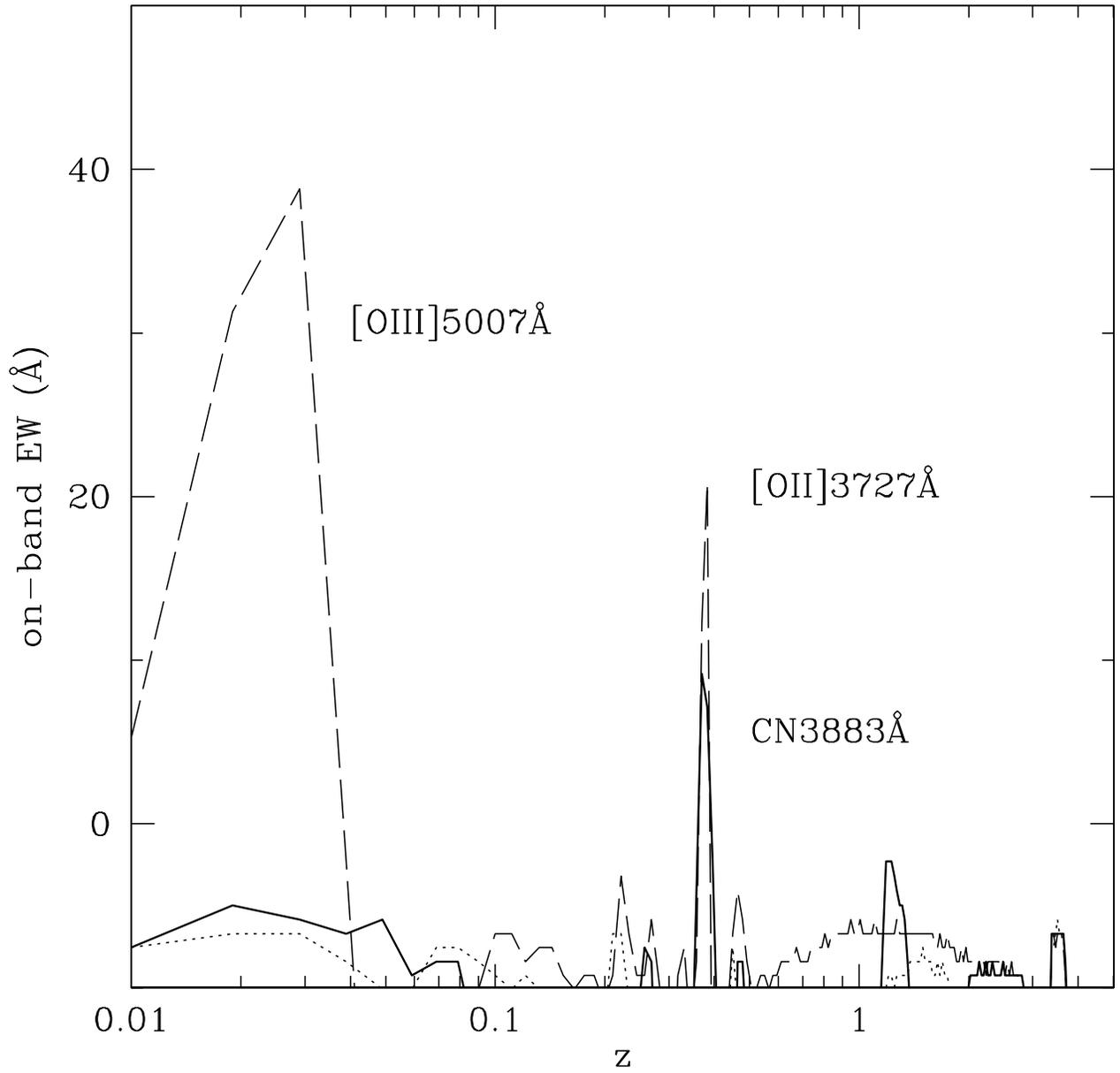}
\caption{ The predicted excess on-band EW for each spectral type 
(E/S0 = solid line, Sbc = dotted line, Im = dashed line) 
as a function of redshift.  The SEDs are from Coleman et al.
 (1980). }
\end{figure}

\clearpage
\begin{figure}
\plotone{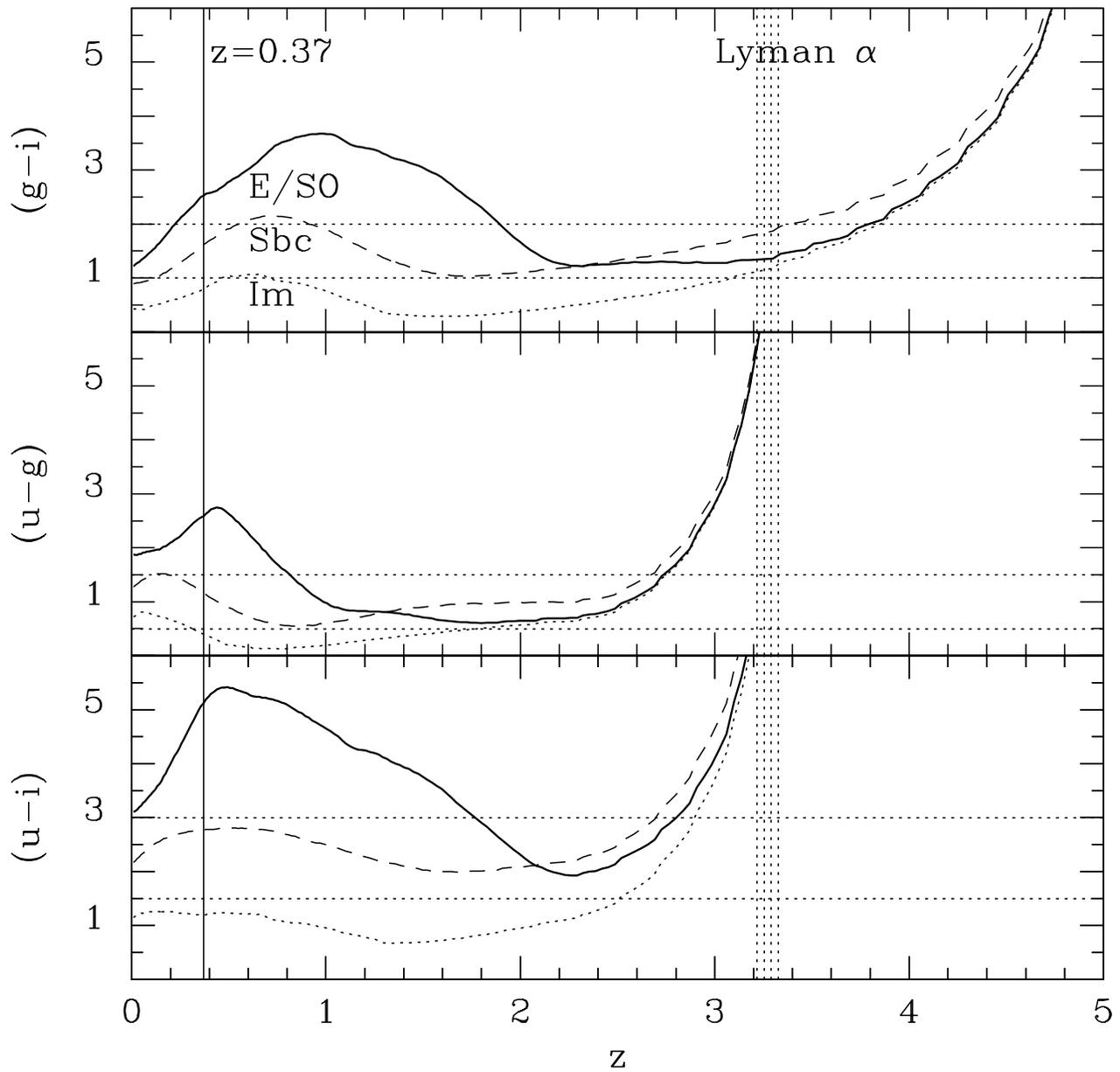}
\caption{ The predicted $u-i$, $u-g$, and $g-i$ colors for each spectral type 
(E/S0, Sbc, Im) as a function of redshift.  The SEDs are from Coleman et al.
(1980). The solid vertical line shows the redshift of the
MS1512.4+3647 cluster. The dashed vertical lines show the redshift range over
which background Lyman $\alpha$ emitters could be detected.}
\end{figure}

\clearpage
\begin{figure}
\plotone{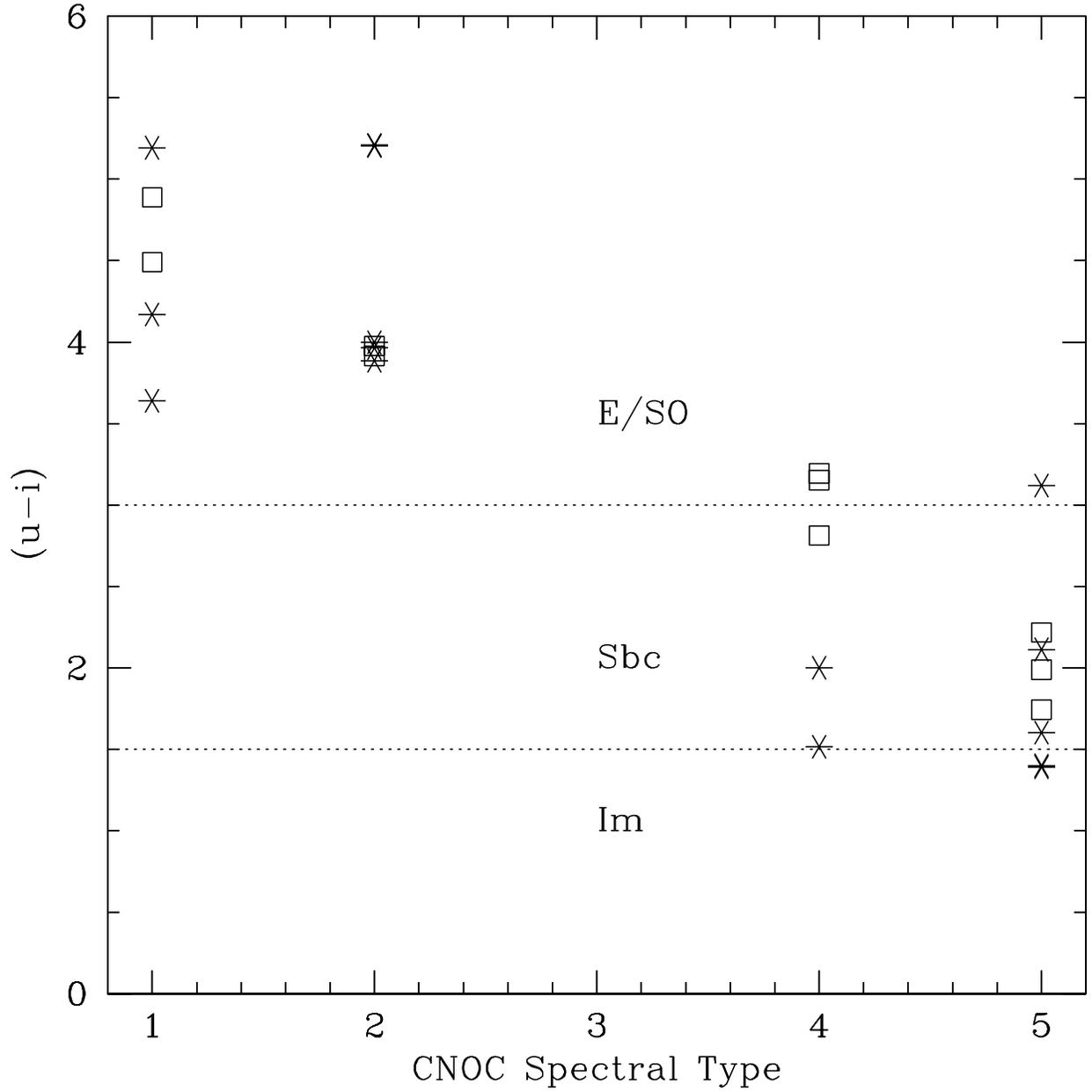}
\caption{ The $u-i$ colors of the 36 cluster galaxies observed by
CNOC vs the CNOC spectral classification (E/S0 =1,2, Spiral=4, Emission-line/
Im = 5). Objects with on-band excesses greater than 3 $\sigma$ are starred. }
\end{figure}

\clearpage
\begin{figure}
\plotone{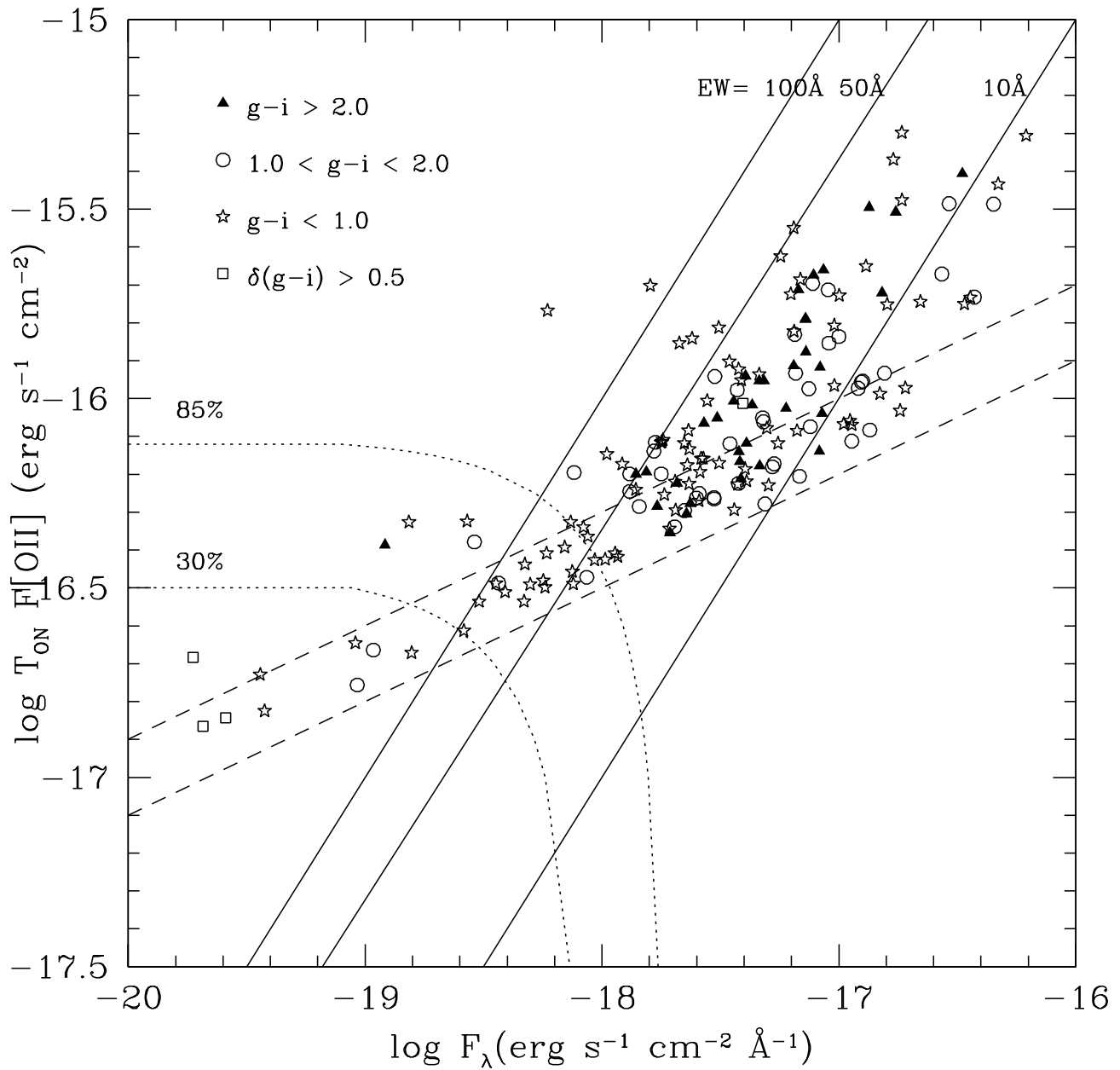}
\caption{ T$_{ON}$F[OII] vs. continuum flux for 3 $\sigma$ on-band excess objects.
The upper dashed line shows the 4 $\sigma$ cutoff limit, and the curves in the
lower left are for 30\% and 85\% completeness. The solid diagonal lines are lines
of constant EW. }
\end{figure}

\clearpage
\begin{figure}
\plotone{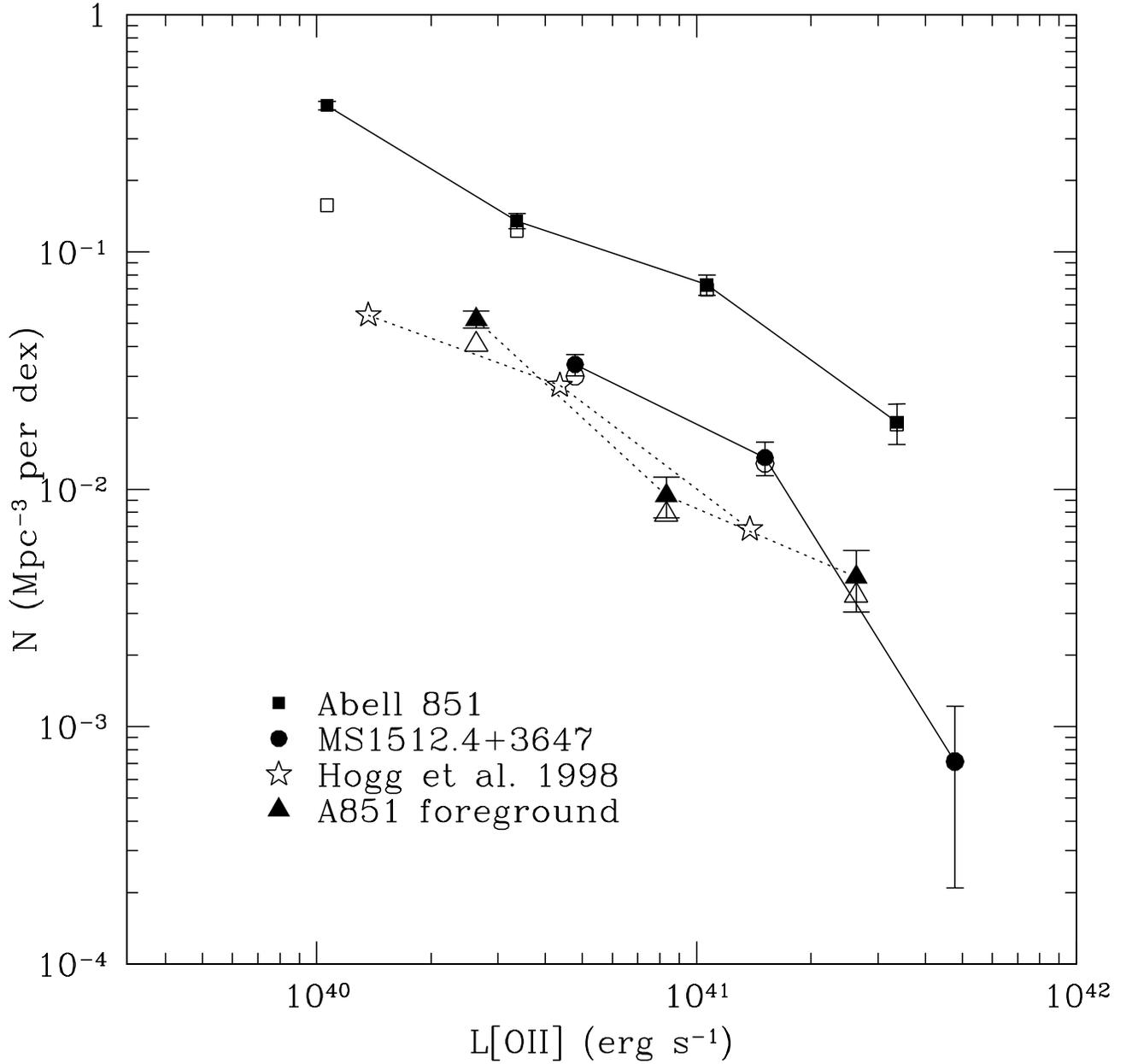}
\caption { The [OII] 3727\AA\ luminosity functions of the MS1512.4+3647 
cluster, Abell 851 cluster and foreground fields 
(Martin et al. 2000), and the field at $z \sim$ 0.4 (Hogg et al. 1998). The
open symbols give the luminosity function prior to the incompleteness
correction.}
\end{figure}

\clearpage
\begin{figure}
\plotone{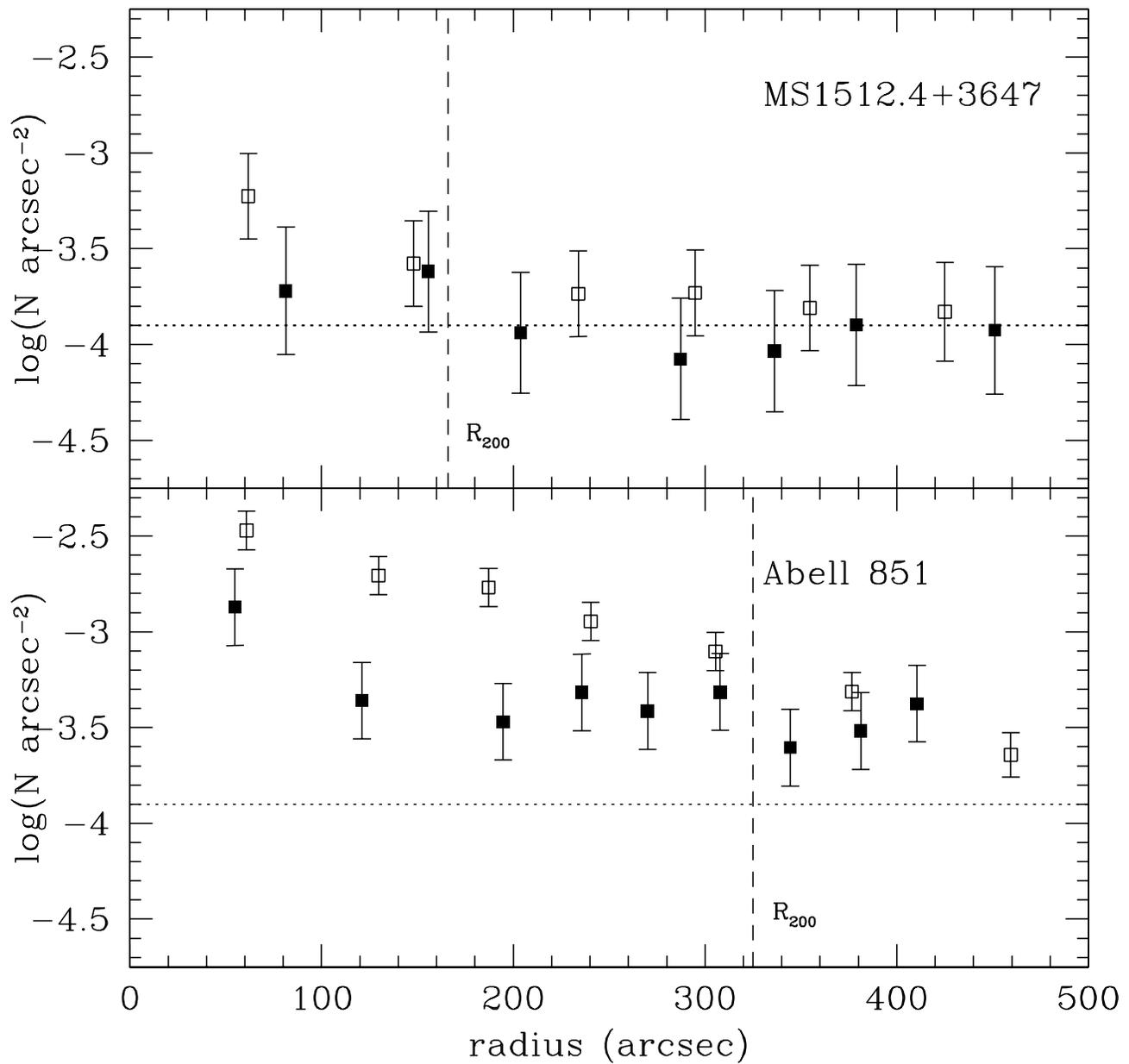}
\caption{ The surface density of [OII] emission-line candidates 
(filled squares) and 
red galaxies with $g-i > 2.0$ (open squares) in
the fields of the MS1512.4+3647 and Abell 851 clusters.  
The dotted line gives the expected surface density
of field [OII] emitters within our on-band (Hogg et al. 1998).  
R$_{200}$ is shown for each cluster.}
\end{figure}

\clearpage
\begin{figure}
\plotone{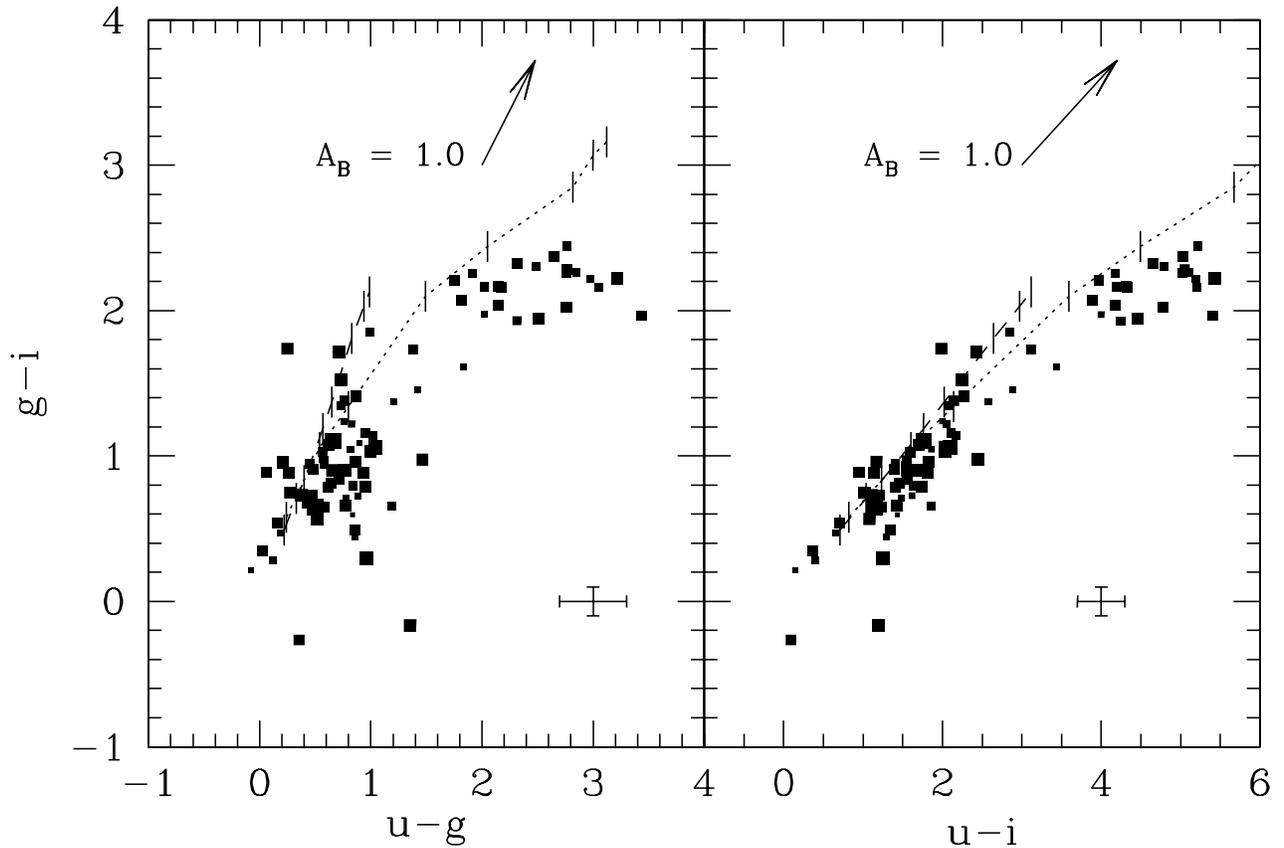}
\caption{ $g-i$ vs $u-g$ (left) and $u-i$ (right) 
for the MS1512.4+3647 on-band excess objects
(filled squares).  The sizes of the points are proportional to log([OII] EW). 
The error bars show the typical photometric errors of the
galaxies at $i=23$.  The model tracks are the predicted colors for a burst star 
formation history (dotted line) and a constant star-formation history 
(dashed line) observed at redshift $z=0.37$ (Bruzual \& Charlot 2000). 
The tick marks on the model tracks are at 10 Myr, 50 Myr, 100 Myr, 500 Myr, 
1 Gyr, 5 Gyr, 10 Gyr and 13 Gyr.  
The models assume an internal extinction of A$_B$ = 1.0, $Z = 1/3 \Zsun$ and
a Salpeter IMF. }
\end{figure}

\clearpage
\begin{figure}
\plotone{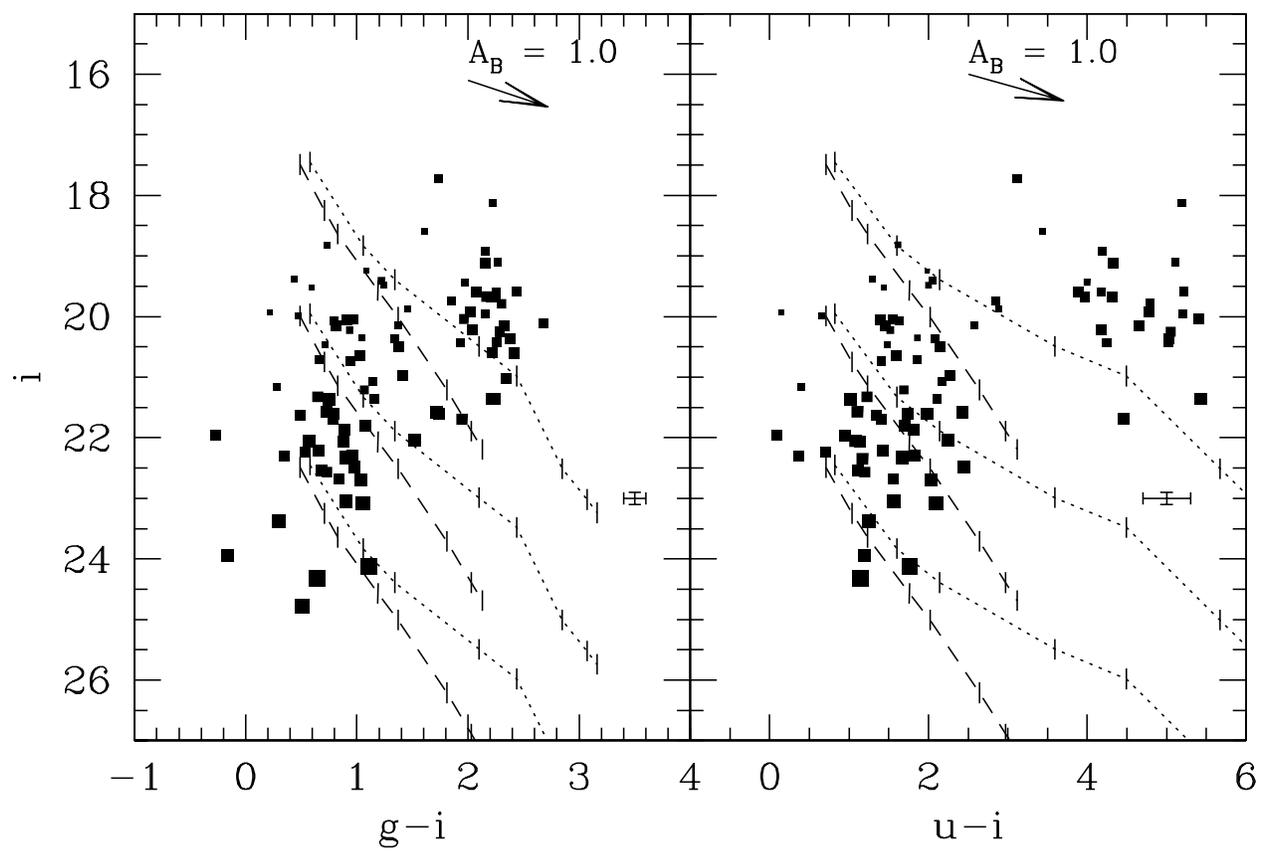}
\caption{ $i$ vs $g-i$ (left) and $u-i$ (right) 
for the MS1512.4+3647 on-band excess objects
(filled squares). The size of the points is proportional to log( [OII] EW).
The error bars shows the typical photometric errors of the galaxies at $i$ = 
23. The models tracks are for constant
 mass (10$^{10}$, 10$^{9}$, and 10$^{8}$ $\Msun$)
and increasing age for a burst star-formation history (dotted line) and a 
constant star-formation history (dashed line; same as Figure 9).  }
\end{figure}

\clearpage
\begin{figure}
\plotone{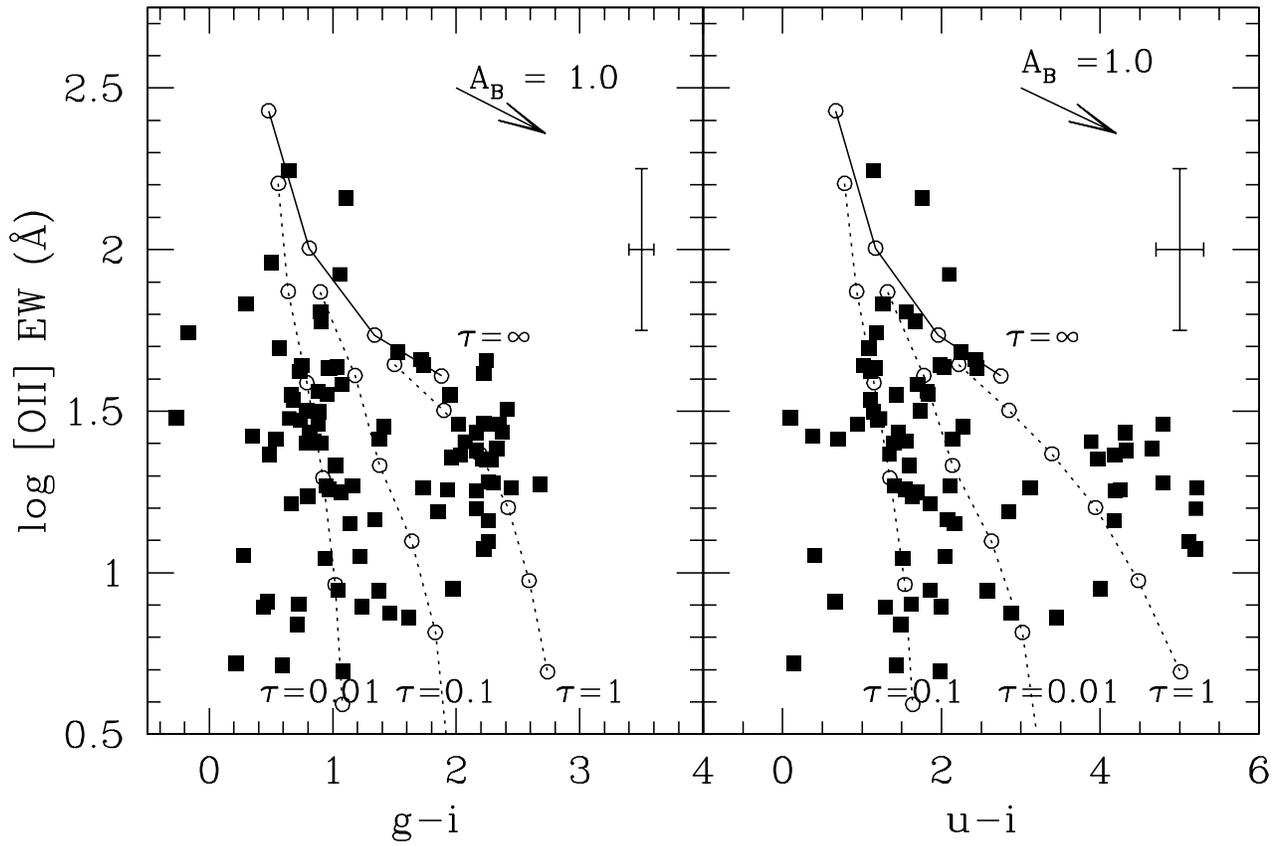}
\caption{ log [OII] EW vs $g-i$ (left) and $u-i$ (right) 
for the MS1512.4+3647 on-band 
excess objects (filled squares). The error bars are for a 4$\sigma$ 
detection at $i= 23.0$. The model tracks are for exponentially decaying star-formation
histories ($\tau$ = 0.01 Gyr, 0.1 Gyr, 1 Gyr, and $\infty$), with 
the same dust and metallicity assumptions as the previous figures. 
Ages of 10 Myr, 100 Myr, 1 Gyr,  and 10 Gyr are shown for $\tau = \infty$ track.
The other tracks have ages 1-6 $\tau$ marked. }
\end{figure}

\clearpage
\begin{figure}
\plotone{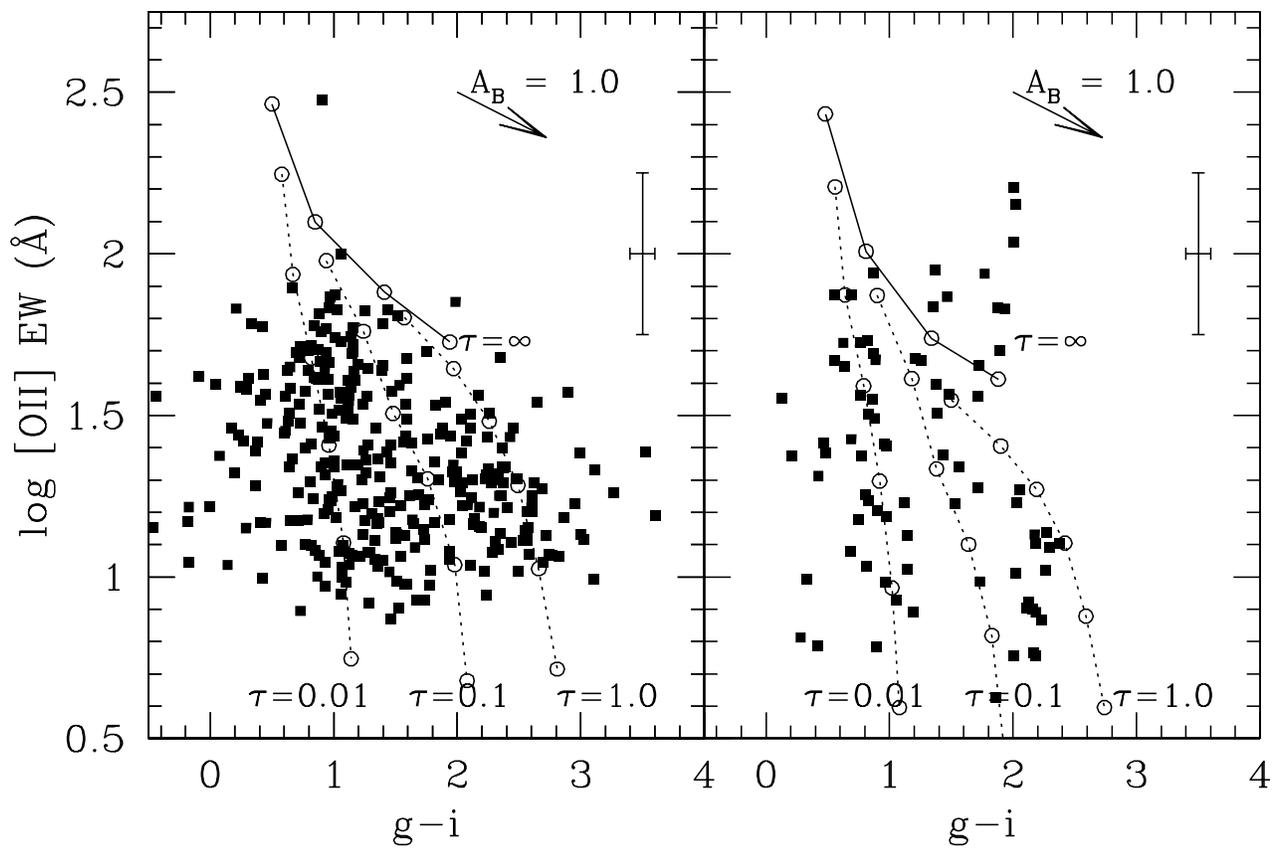}
\caption{ log [OII] EW vs $g-i$ for Abell 851 on-band excess cluster objects
(left) and off-band excess foreground objects (right).
The model tracks are same as in Figure 11. 
}
\end{figure}

\clearpage
\begin{figure}
\plotone{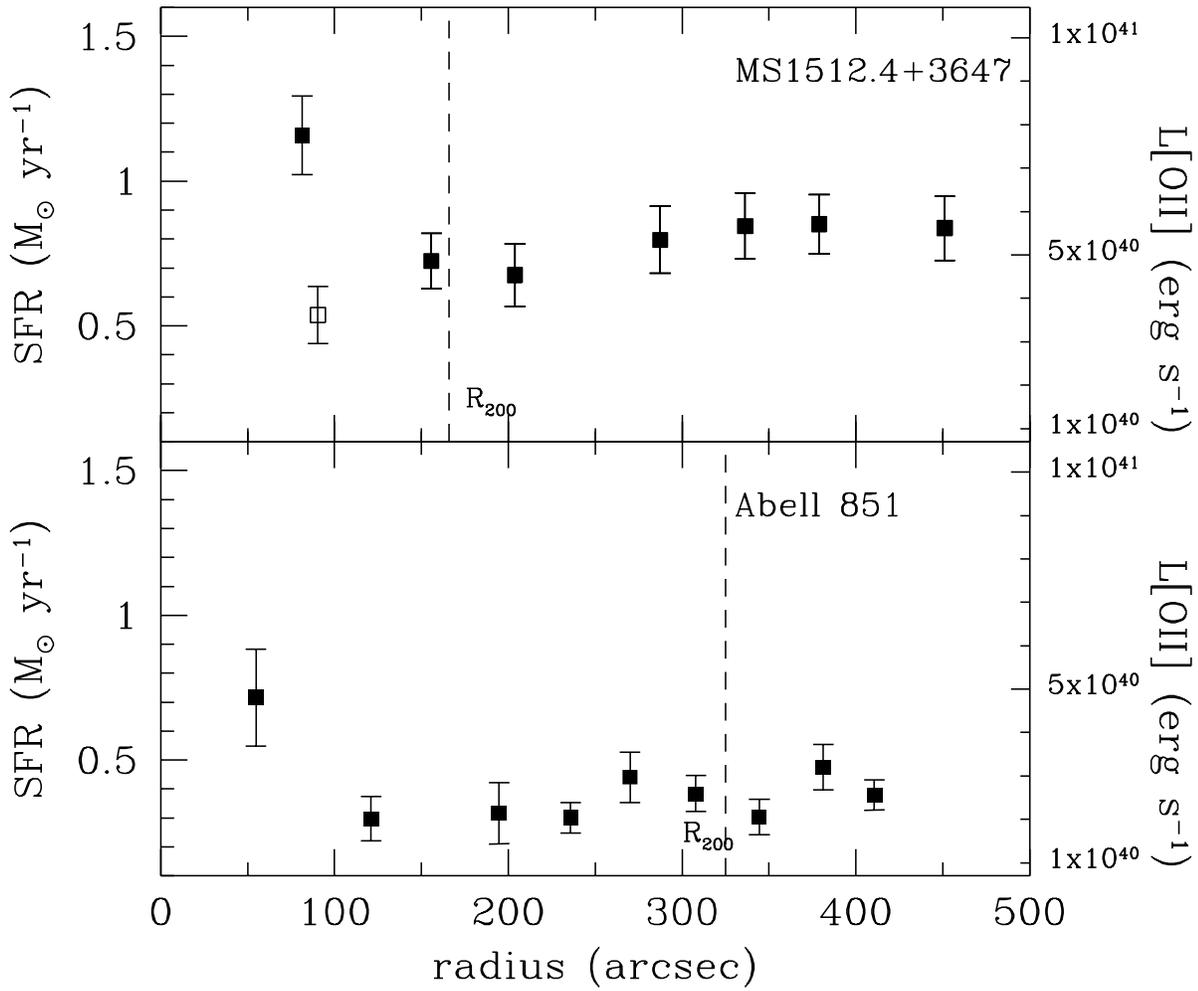}
\caption{Average star-formation rate and [OII] luminosity of the MS1512.4+3647 
and Abell 851 emission-line candidates as a function of cluster radius. 
The open symbol (top) is the average SFR in the inner 275 kpc excluding MS1512.4+3647's
central cluster galaxy. (Note that the Abell 851 sample has
a lower limiting [OII] flux than the MS1512.4+3647 sample, therefore the 
average SFR rate for the detected Abell 851 sample is lower).  }
\end{figure}

\end{document}